\documentclass{article}

\usepackage{arxiv}

\usepackage[utf8]{inputenc}
\usepackage[T1]{fontenc}
\usepackage{textcomp}
\usepackage{amsmath,amssymb,amsfonts}
\usepackage{graphicx}
\usepackage{booktabs}
\usepackage{multirow}
\usepackage{array}
\usepackage{caption}
\usepackage{microtype}
\usepackage{newunicodechar}
\usepackage{xurl}
\usepackage[hidelinks]{hyperref}

\newunicodechar{−}{\ensuremath{-}}
\newunicodechar{×}{\ensuremath{\times}}
\newunicodechar{°}{\ensuremath{^{\circ}}}
\newunicodechar{±}{\ensuremath{\pm}}
\newunicodechar{ₓ}{\textsubscript{x}}
\newunicodechar{₂}{\textsubscript{2}}
\newunicodechar{₃}{\textsubscript{3}}
\newunicodechar{₄}{\textsubscript{4}}
\newunicodechar{²}{\textsuperscript{2}}
\newunicodechar{⁻}{\textsuperscript{\ensuremath{-}}}
\newunicodechar{⁶}{\textsuperscript{6}}
\newunicodechar{λ}{\ensuremath{\lambda}}
\newunicodechar{κ}{\ensuremath{\kappa}}
\newunicodechar{σ}{\ensuremath{\sigma}}
\newunicodechar{τ}{\ensuremath{\tau}}
\newunicodechar{Δ}{\ensuremath{\Delta}}
\newunicodechar{Ω}{\ensuremath{\Omega}}

\renewcommand{\headeright}{}
\renewcommand{\undertitle}{}
\renewcommand{\shorttitle}{Atmospheric chemistry-climate modeling with an AI foundation model}

\title{Atmospheric Chemistry-Climate Modeling with\\ An AI Foundation Model}

\author{
  James Weber\thanks{These authors contributed equally to this work.} \\
  Department of Meteorology\\
  University of Reading, Reading, UK\\
  \texttt{j.m.weber@reading.ac.uk} \\
  \And
  Makoto Kelp\footnotemark[1] \\
  Department of Atmospheric Sciences\\
  University of Utah, Salt Lake City, UT, USA\\
  \texttt{makoto.kelp@utah.edu} \\
}

\date{}

\begin{document}
\maketitle
{\renewcommand{\thefootnote}{}\footnotetext{This manuscript as presented here is a non-peer reviewed preprint currently submitted to a scientific journal.}}
\begin{abstract}
Atmospheric composition and chemistry-climate interactions are central to predicting future climate change. Yet simulating chemical and aerosol processes is typically the most expensive component of an atmospheric model and is often highly simplified or prescribed entirely. AI weather forecasting foundation models have shown skill rivalling process-based models and are increasingly fine-tuned for new atmospheric tasks. Here, we present the first foundation model to capture interactions between emissions, atmospheric composition, and radiation. We fine-tune the Aurora Earth system foundation model on simulations from the chemistry-climate model UKCA. Our model reproduces daily meteorology, composition, aerosol, and radiation fields of the coupled system with 23 of 27 predicted fields achieving R² above 0.85 in both the troposphere and stratosphere. It reproduces ozone, CO, and temperature with annual mean biases below 4\% and the downwelling shortwave and longwave fluxes to within 6\%. Our model remains stable in autoregressive simulation over 50 years while running \textasciitilde100× faster than UKCA. By training on paired emissions scenarios, our model learns responses to perturbations by capturing tropospheric ozone's response to changes in NO and CO emissions and surface radiation's response to aerosol. We further show that incorporating MLS satellite ozone retrievals into training corrects biases in the underlying UKCA fields, reducing zonal mean ozone error against collocated observations by 23\% between 250 and 20 hPa. Our approach does not suggest process-based modeling can be replaced but demonstrates that a foundation model can be adapted to respond to forcings and learn from observations at a relatively small computational cost.
\end{abstract}

\section{Introduction}

Global chemistry-climate models (CCMs) simulate the interactions between gas phase composition, aerosols, radiation, clouds, and meteorology. They quantify how changes in emissions affect atmospheric chemistry, radiative forcing, and climate feedbacks (e.g., Thornhill et al., 2021a, 2021b). This data informs air quality and climate mitigation strategies by assessing the benefits and trade-offs of emission reductions (e.g., Allen et al., 2021; Turnock et al., 2023; Weber et al., 2026, 2024). The response of the Earth's radiative balance and air quality to a forcing arises from multiple interacting processes. For example, increases in nitrogen oxides (NO\textsubscript{x}) emissions drive ozone production (warming) but reduce methane lifetime and increase nitrate aerosol (both cooling), while also worsening surface air quality (Thornhill et al., 2021b). Capturing their net effect requires explicit simulation of composition and its interaction with meteorology, which is computationally demanding. CCMs must integrate stiff chemical systems across species with widely differing lifetimes, transport large numbers of tracers, and calculate their radiative effects (Archibald et al., 2020). Simulating chemical reactions can take up 50-90\% of a model's runtime (Eastham et al., 2018; Long et al., 2015), limiting chemical complexity and the number of simulations available for scenario analysis.

Artificial intelligence (AI) modeling offers several approaches to reducing the cost of atmospheric simulation. In atmospheric chemistry, neural networks have been used to replace expensive numerical chemical solvers while retaining the surrounding transport and physical parameterizations (Kelp et al., 2022, 2020; Xia et al., 2025), and Gaussian processes have emulated the surface temperature response to aerosol emission changes (Dewey et al., 2025). Conservation properties can also be incorporated directly into AI architectures to preserve physical relationships (Sturm and Wexler, 2022, 2020). However, learned components can develop biases or instabilities when coupled to other processes in their parent models, including through operator splitting (Brenowitz et al., 2020; Keller and Evans, 2019; Kelp et al., 2018; Rasp, 2020). AI models can also predict the full atmospheric state. Pangu-Weather and GraphCast outperform operational numerical weather forecasts across many variables and lead times at substantially lower cost once trained (Bi et al., 2023; Lam et al., 2023), while NeuralGCM couples a numerical dynamical core with learned parameterizations and the AI2 Climate Emulator predicts atmospheric evolution directly, enabling simulations over decades (Kochkov et al., 2024; Watt-Meyer et al., 2025). However, these weather and climate models do not include interactive chemistry, aerosols, or their radiative effects. Extending these approaches to chemistry-climate modeling requires accounting for the different timescales and feedbacks introduced by reactive gases and aerosols.

AI foundation models offer a path to develop these capabilities by adapting (or ``fine-tuning'') models already trained on large atmospheric datasets to new variables, resolutions, and applications. ClimaX, pretrained on climate simulations, has been adapted to weather forecasting and climate projection (Nguyen et al., 2023). NASA-IBM's Prithvi WxC extends the approach to forecasting, downscaling, and gravity wave parameterization using a model pretrained on atmospheric reanalysis (Schmude et al., 2024). Microsoft's Aurora enables weather, air pollution, and ocean wave forecasting through specialized versions of the same pretrained model (Bodnar et al., 2025). Such models offer a mature starting point for jointly predicting composition, meteorology, and radiation without embedding separate emulators in numerical parent models. However, successful adaptation does not establish that a model represents the physical relationships needed for a task. Hu et al. (2026) found that Aurora's air pollution model reproduced some ozone responses to NO\textsubscript{x} perturbations but did not consistently represent anomalies such as wildfire smoke transport, underscoring the importance of how AI models are fine-tuned and evaluated (Kelp et al., 2026). For CCMs, agreement with reference concentration fields must be complemented by tests of chemical and radiative responses to changing emissions.

Accurate model emulation does not ensure agreement with the observed atmosphere. Even detailed CCMs retain biases from uncertain emissions, reaction rates, and parameterized processes (e.g., Sporre et al., 2020; Squire et al., 2015; Weber et al., 2021; Wild et al., 2020; Wild and Ryan, 2026). Models trained on simulations or reanalysis can inherit these biases (Inness et al., 2019). Observations guide calibration, but translating observed discrepancies into corrections to coupled process representations remains difficult (Hourdin et al., 2017; Schneider et al., 2017). AI methods have been used to correct surface ozone biases in CCM simulations and chemical reanalysis, but these corrections operate on output fields without changing the underlying model evolution (Liu et al., 2025; Miyazaki et al., 2025). Conventional chemical data assimilation adjusts concentrations and emissions within existing process formulations (Miyazaki et al., 2020), while recent approaches perform assimilation in the learned latent space of an AI atmosphere model (Fan et al., 2026).

Observational fine-tuning offers an alternative that does not require a dedicated data assimilation framework. It adjusts a foundation model's learned evolution during training, allowing measurements to influence subsequent predictions without a separate assimilation step during inference. For example, fine-tuning Prithvi WxC on satellite precipitation estimates improved forecast accuracy against independent observations (Pfreundschuh et al., 2026). Extending this approach to atmospheric chemistry is complicated by the generally sparser coverage of chemical measurements compared with meteorological observations (Bocquet et al., 2015). Although ozone is the most measured trace gas in the atmosphere, measurement networks provide uneven geographic coverage and instruments differ in their sampling frequency and vertical sensitivity (Kelp et al., 2025; Van Malderen et al., 2025). To our knowledge, direct chemical observations have not yet been used to fine-tune a foundation model for global chemistry-climate simulation.

Here, we present the first foundation model adapted for interactive chemistry-climate simulation. We fine-tune the Aurora Earth system foundation model on simulations from the United Kingdom Chemistry and Aerosols (UKCA) CCM to predict tropospheric and stratospheric composition alongside interacting meteorology, clouds, and radiative fluxes. We first evaluate daily prediction accuracy, computational cost, and stability in free-running simulations of up to 50 years. We then train on paired emission scenarios and test the responses of ozone to NO\textsubscript{x} and carbon monoxide (CO) emissions and of surface radiation to black carbon (BC) aerosol emissions. Finally, we incorporate Aura Microwave Limb Sounder (MLS) ozone retrievals into training, without supplying observations during inference, and evaluate the model against held-out retrievals. This work provides a basis for developing AI Earth system models that learn from both simulations and observations to represent feedbacks between atmospheric composition and climate.

\section{Materials and Methods}

Our framework adapts the pretrained Aurora model (Bodnar et al., 2025) to predict the daily evolution of UKCA output (Figure 1). We retain Aurora's core architecture and introduce variable transformations and training objectives that account for the concentration ranges, timescales, and conservation properties of atmospheric chemistry. This framework yields three model variants. Aurora\textsubscript{UKCA} is trained on the UKCA control simulation and evaluated for baseline performance and stability in autoregressive simulations spanning one to fifty years. Further fine-tuning on UKCA simulations with perturbed emissions produces Aurora\textsubscript{UKCA,emiss}, which is trained to reproduce chemical and radiative responses to emission changes. A separate branch incorporates satellite ozone observations alongside nudged UKCA output to produce Aurora\textsubscript{obs} and assess whether observational training reduces inherited UKCA biases. Below, we describe the datasets, model architecture, and training procedure before detailing the evaluation and additional fine-tuning experiments.

\begin{figure}[htbp]
\centering
\includegraphics[width=\textwidth,height=0.82\textheight,keepaspectratio]{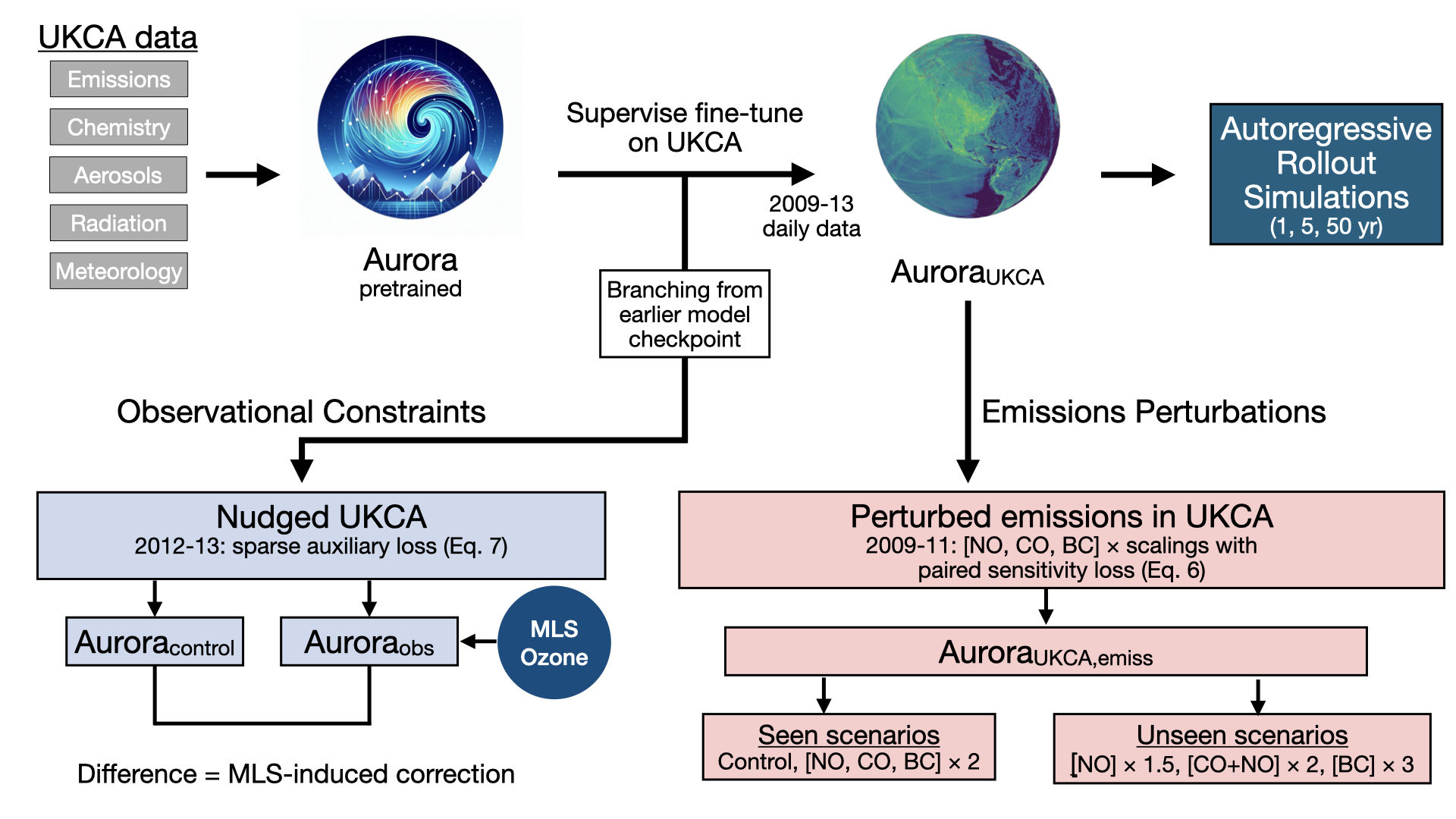}
\caption{\textbf{Adapting the Aurora Earth system foundation model into a data-driven UKCA chemistry-climate model.} Aurora ingests two input time steps on 20 pressure levels (1000-0.1 hPa), upsampled from UKCA's 1.25° × 1.875° grid to 0.5° with chemical concentrations passed through a learnable log-linear transform. We (1) fine-tune on free-running UKCA output and evaluate single step accuracy and 1-, 5-, and 50-year rollouts; (2) further fine-tune on UKCA simulations with scaled NO, CO, and BC emissions using a paired sensitivity loss; and (3) add a sparse auxiliary loss against Aura MLS ozone retrievals between 250 and 20 hPa.}
\label{fig:1}
\end{figure}

\subsection{UKCA model simulations}

We fine-tune Aurora on output from UKCA simulations (Table 1). UKCA is the atmospheric composition component of the United Kingdom Earth System Model (UKESM) (Sellar et al., 2019). UKCA simulates interactive tropospheric and stratospheric chemistry and aerosol processes, photolysis, and dry and wet deposition. All simulations are run at a horizontal resolution of 1.25° x 1.875° with 85 vertical levels up to 85~km (Walters et al., 2019), and use the GLOMAP-mode aerosol scheme, which simulates sulfate, sea salt, BC, organic carbon (OC), and dust (Mulcahy et al., 2023) and allows aerosol to interact with radiation. Specifically, we run UKCA in the UKESM atmosphere-only (AMIP) configuration at Unified Model (UM) version 12.0 with Strat-Trop chemistry.

The free-running simulations use prescribed 12-month climatologies of sea surface temperature (SST), sea ice, land cover, leaf area index, mean canopy height, and oceanic aqueous dimethyl sulfide (DMS) and chlorophyll concentrations. These climatologies are derived from a coupled UKESM1.0 simulation spanning 2005-2014. The use of prescribed SSTs reduces the noise which would occur with a coupled ocean model configuration. The simulations use prescribed 12-month CMIP6 2014 timeslice anthropogenic and biomass burning emissions (Hoesly et al., 2018), oceanic VOC emissions from the POET dataset, biogenic emissions from the MEGAN-MACC 2001-2010 climatology (Guenther et al., 2012) and interactive isoprene and monoterpene emissions from the iBVOC module (Pacifico et al., 2011). Prescribed surface concentrations of CO\textsubscript{2}, methane, N\textsubscript{2}O, CFC11, and CFC12 and other trace halocarbons are taken from the end of the historical period. The ``Nudged UKCA'' run uses observed SSTs and sea ice and historical emissions while meteorology (temperature and horizontal winds) is nudged to ERA5 (Telford et al., 2013).

\begin{table}[htbp]
\centering
\caption{UKCA simulations used to fine-tune the Aurora foundation model.\textsuperscript{a}}
\label{tab:sims}
\small
\begin{tabular}{@{}lllp{5.6cm}@{}}
\toprule
\textbf{Simulation} & \textbf{Purpose} & \textbf{Period} & \textbf{Details}\textsuperscript{b} \\
\midrule
Base & Train and Test & \begin{tabular}[t]{@{}l@{}}2009-2013 (Train)\\2014 (Test)\end{tabular} & 2014 emissions \\
\midrule
NO\_0.5x & \multirow{7}{*}{Train and Test} & \multirow{7}{*}{\begin{tabular}{@{}l@{}}2009-2010 (Train)\\2011-2014 (Test)\end{tabular}} & Surface NO emissions $\times$ 0.5 \\
NO\_2x & & & Surface NO emissions $\times$ 2 \\
CO\_0.5x & & & CO emissions $\times$ 0.5 \\
CO\_2x & & & CO emissions $\times$ 2 \\
BC\_0.5x & & & BC emissions $\times$ 0.5 \\
BC\_2x & & & BC emissions $\times$ 2 \\
BC\_5x & & & BC emissions $\times$ 5 \\
\midrule
NO\_1.5x & \multirow{4}{*}{Test only} & \multirow{4}{*}{2009-2011} & Surface NO emissions $\times$ 1.5 \\
CO\_1.5x & & & Surface CO emissions $\times$ 1.5 \\
NO\_CO\_2x & & & Surface NO + CO emissions $\times$ 2 \\
BC\_3x & & & BC emissions $\times$ 3 \\
\midrule
Nudged UKCA & Train and Test & 2009-2013 & Nudged to ERA5 fields for u, v and temperature down to \textasciitilde1200 m. Historical emissions \\
\bottomrule
\end{tabular}

\vspace{4pt}
\begin{minipage}{\textwidth}\footnotesize
\textsuperscript{a}Daily mean output on UKCA's native 1.25° latitude by 1.875° longitude grid at 20 pressure levels: 1000, 925, 850, 700, 600, 500, 400, 300, 250, 200, 150, 100, 70, 50, 30, 20, 10, 5, 1, 0.1 hPa.\\
\textsuperscript{b}All anthropogenic and biomass burning emissions are from 2014 unless otherwise stated.
\end{minipage}
\end{table}

\subsubsection{Physical Variables}

Surface variables comprise 2-m temperature, 10-m zonal and meridional winds, surface pressure, and mean sea level pressure. At each of the 20 pressure levels, we track zonal and meridional winds, temperature, geopotential height, specific humidity, and upward and downward shortwave (SW) and longwave (LW) radiative fluxes.

We reduce the 3D cloud fraction field to a binary low-cloud indicator to reduce complexity. The low cloud variable is set to 1 where cloud fractions summed over the four lowest atmospheric levels (1000-700 hPa) exceed 1.0, and 0 otherwise. A high-cloud indicator is also defined and trained, but UKCA reports cloud fraction as missing at 30 hPa and above. As a result, we exclude it from the model's outputs.

\subsubsection{Composition Variables}

At each pressure level, we represent 15 gas phase species and chemical families. These comprise oxidants and radicals (ozone, OH, HO\textsubscript{2}), carbon species (CO, methane, isoprene, HCHO), and reactive nitrogen species (NO\textsubscript{x}=NO+NO\textsubscript{2}, HNO\textsubscript{3}, PAN). Halogen chemistry is represented by HCl, ClO\textsubscript{x} (=Cl+ClO), Cl\_precursor (=CFC11+CFC12), ClONO\textsubscript{2}, and MeBr. We also include SO\textsubscript{2}, sulfate, and BC in the soluble aerosol accumulation mode, giving 18 composition variables. All composition variables are expressed as mass mixing ratios. Surface emission fields comprise NO, CO, isoprene, HCHO, BC aerosol, and OC aerosol. This selection represents a subset of UKCA's gas and aerosol composition; extending the species and aerosol modes represented is a direction for future development.

\subsection{Aurora Model Architecture}

We fine-tune the Aurora Earth system foundation model pretrained by Bodnar et al. (2025). The original model is built on ERA5 reanalysis, CMIP6 climate simulations, and GFS operational analyses largely for medium-range weather forecasting. The model architecture comprises three main components: (1) an encoder that compresses and embeds each gridded surface and atmospheric variable into patches of ``tokens'' (i.e., a vector representation of that patch), and aggregates pressure levels into a small number of latent levels; (2) a 3D Swin Transformer U-Net backbone that maps the current state to a future state conditioned on lead time; and (3) a decoder that reconstructs gridded fields from the processed latent tokens (Figure 1).

We make minimal changes to Aurora's base architecture. UKCA's native model grid uses 1.25° by 1.875° resolution. Before patch embedding, we bilinearly upsample this output to 0.5° resolution to closely align our data with the resolution regime at which Aurora's pretrained weights were learned (0.25° resolution). We select 0.5° resolution as a tractable compromise as it is closer to Aurora's native training resolution than UKCA's native grid while remaining within the memory budget of the available hardware. The forecast lead time for our fine-tuning is set at 24 hours, matching our training data's daily temporal resolution rather than Aurora's native 6-hour time step.

Representing atmospheric chemistry within this architecture requires adding physical and statistical properties that the pretrained Aurora's native meteorological variables do not share. Aurora was designed to predict fields that can take both positive and negative values and vary over a comparatively narrow dynamic range. Trace gas mixing ratios hold neither of these properties. For example, a species such as OH can vary by four or more orders of magnitude between polluted and background air. Every chemistry variable must also remain non-negative, a constraint that standard z-score normalization does not enforce. We address the dynamic-range problem by constructing an input transform (Eq. 1) that blends a linear and a logarithmic representation of concentration, such that each species can be weighted toward whichever functional form better captures its concentration distribution:

\begin{equation}
f(z) = w_{1}\,\min\left(\max(z,0),\,2.5\right) + w_{2}\,\frac{\log\left(\max(z,10^{-4})\right) - \log\left(10^{-4}\right)}{-\log\left(10^{-4}\right)}
\end{equation}

We transform each chemistry input value \emph{z} before the encoder through a learnable per-variable weighted combination of two terms: a term linearly bounded between 0 and 2.5 and a rescaled log term. The value \emph{z} is normalized per species and pressure level by half the time-mean of the instantaneous spatial maximum concentration, such that \emph{z}=2 corresponds to that typical maximum. Bounding this term at 2.5 covers roughly 25\% above that maximum while keeping extreme outliers from dominating it. The weights \emph{w1} and \emph{w2} are set equal in value at the start and are then optimized during training, so that the model can shift each weight toward whichever representation best matches each species' distribution. Because this transform acts on the encoder inputs rather than the decoder outputs, it does not by itself guarantee non-negative predictions. Therefore, we clip predicted concentrations at zero before they are evaluated or fed back autoregressively.

We represent cloud fraction differently from the remaining atmospheric variables. Rather than predicting a continuous 3D cloud fraction field, we predict the binary low-cloud variable defined in Section 2.1.1. This field is excluded from the log-linear transform above and is trained against its binary target.

\subsection{Loss Functions}

A loss function is the quantity the AI model is trained to minimize, i.e., a numerical measure of how far a prediction differs from the true UKCA state. Loss functions are one of the primary mechanisms by which Aurora's weights are adapted for a new task, more so than any model architectural change. As a result, loss functions become the primary avenue through which we convert a weather forecasting model into a CCM and constitute most of the changes we make beyond data curation and the base architecture.

For the primary chemistry loss, we normalize by the root mean square size of each variable's day-to-day change computed over the full UKCA training record (Watt-Meyer et al., 2023). This loss was originally deployed in the AI2 Climate Emulator for general atmospheric state variables, but we extend it here to trace gas species (Eq. 2). Bodnar et al.'s pretraining loss normalizes each variable by its climatological standard deviation, which is a typical design choice for meteorological fields whose variability is comparatively stable year to year. This approach is not well suited to atmospheric chemistry, because two species can share similar climatological variance while their typical day-to-day concentration change differs by orders of magnitude. As a result, a longer rollout would not receive comparable loss across species. Furthermore, atmospheric lifetimes for the predicted chemical species can range over many orders of magnitude. For example, methane has a tropospheric lifetime on the order of a decade, while OH is destroyed within about a second of formation. An unnormalized loss would therefore be dominated by whichever species has the largest absolute day-to-day variability, rather than treating every species on comparable terms. Past machine learning studies of atmospheric chemistry have found that error accumulation is driven more by slower evolving chemical modes (Keller and Evans, 2019; Kelp et al., 2020).

For variable \emph{v} at pressure level \emph{l}, we define a `tendency' as \(x(t + 1,\ v,\ l) - x(t,\ v,\ l)\). We use the root mean square size of this increment over the full training period as a per-variable, per-pressure level scale factor:
\begin{equation}
\sigma(v,l) = \sqrt{\frac{1}{T}\sum_{t=1}^{T}\left(x(t+1,v,l) - x(t,v,l)\right)^{2}}
\end{equation}

and the normalized loss becomes:

\begin{equation}
L(v,l) = \frac{1}{N}\sum_{i=1}^{N}\frac{\left|\mathrm{pred}_{i}(v,l) - \mathrm{target}_{i}(v,l)\right|}{\sigma(v,l)}
\end{equation}

where \emph{T} is the number of day-to-day pairs in the training period and \emph{N} is the number of grid cells at level \emph{l}, indexed by \emph{i}. These loss terms allow fast- and slow-reacting species to contribute comparably to training.

We apply additional weighting terms on top of this base loss. A factor of 2.0 weight on the four lowest pressure levels (1000-700 hPa) reflects their relevance to surface air quality and prevents their errors from being diluted when averaged across all 20 levels. Grid cells whose target mass mixing ratio falls below 10\textsuperscript{-15} are excluded from the chemistry loss with a higher threshold of 10\textsuperscript{-13} for isoprene, MeBr, Cl\_precursor, ClO\textsubscript{x}, and PAN, because concentrations below these thresholds mostly reflect numerical noise. Predictions of PAN, isoprene, and the soluble aerosol modes are restricted to pressures below 150 hPa, and MeBr to pressures below 10 hPa, reflecting concentrations that are negligibly small above these levels in both the training targets and the model's predictions.

We apply a second, complementary per-species weighting derived from the temporal autocorrelation of each species' global mean concentration in the training record, which acts as a moderate empirical reweighting between species (Text S1).

Finally, we enforce approximate mass conservation for long-lived species through a dedicated penalty term. AI weather forecasting models usually have no analogous expectation that global temperature or humidity should be conserved across time steps. In contrast, a long-lived chemical tracer's global burden should evolve only through known emission and loss processes. Therefore, we penalize deviations between the predicted and target spatial-mean concentration directly, as a proxy for burden:

\begin{equation}
L_{\mathrm{cons}} = \frac{1}{M}\sum_{v=1}^{M}\frac{\left|\underline{\mathrm{pred}}(v) - \underline{\mathrm{target}}(v)\right|}{\left|\underline{\mathrm{target}}(v)\right|}
\end{equation}

where \emph{M} is the number of species with an estimated lifetime of 180 days or longer and each term compares the predicted and target spatial mean concentration.

Composition, meteorology, radiation, clouds, and the six emission fields all receive a loss during training. All but the emissions are also fed back autoregressively (Section 2.5), whereas emissions are prescribed from UKCA at every time step regardless of what the model predicts. Only the surface meteorological fields (which act as externally driven boundary conditions) and the two static fields (land-sea mask and surface geopotential height) receive no loss.

Including emissions in the loss (even though their predictions are discarded every time step) departs from previous demonstrations of fine-tuning Aurora on CAMS reanalysis air pollution data (Bodnar et al., 2025). That model uses fixed monthly-average anthropogenic emissions as a static encoder input. It is one reason Aurora could not accurately predict wildfire persistence and smoke transport (Hu et al., 2026). However, we find that predicting emissions is necessary. Without a loss term on emissions, the gradient contains no term dependent on how emissions are encoded, so fine-tuning constrains neither that encoding nor any relationship between emissions and the other variables. Predicting emissions as an output ties their encoding to the same shared weights that encode composition, meteorology, and radiation, so these relationships are jointly optimized.

Combining every loss term above, the total training loss is:

\begin{equation}
L = L_{\mathrm{atm}} + \lambda_{\mathrm{cons}}L_{\mathrm{cons}} + \lambda_{\mathrm{emit}}L_{\mathrm{emit}} + \lambda_{\mathrm{cloud}}L_{\mathrm{cloud}}
\end{equation}

where L\textsubscript{atm} is the combined chemistry, meteorology, and radiation loss from Eq. 3, with the level and lifetime weights already included. L\textsubscript{cons} is the conservation loss from Eq. 4, weighted by λ\textsubscript{cons} = 0.05. L\textsubscript{emit} and L\textsubscript{cloud} are linear mean absolute error terms on the emission and cloud fields, computed the same way as Eq. 3 but without the σ normalization, as they are 2D fields rather than vertically resolved species. λ\textsubscript{emit} and λ\textsubscript{cloud} are both 1.0.

\subsection{Training Procedure}

We fine-tune the model by optimizing every loss term across a single time step. We then apply the fine-tuned model autoregressively during evaluation, feeding each prediction back in as an input for the next time step. We compare this single-step loss approach against alternatives that optimize directly over multiple time steps during training, including recurrent memory architectures and losses computed over short autoregressive windows. We find that these alternatives are less accurate and become beholden to the time horizon we select, because a model optimized over a fixed multi-step window has no guarantee of generalizing to a different rollout length. Single-step training with autoregressive evaluation avoids this dependence. This design also mirrors process-based CCMs, which do not carry an extended time history either. Retaining such a history would be prohibitively computationally and memory expensive at the resolution and species count considered here and is not how conventional numerical solvers integrate a system forward in time.

Autoregressive rollout introduces ``exposure bias''. Exposure bias is the mismatch between the training context, drawn directly from the UKCA simulation, and the self-generated context Aurora encounters once it feeds its own predictions back as input. We mitigate this by injecting multiplicative noise with standard deviation 0.05 into the input fields at each training step. Our evaluation rollouts run for a full year of daily steps, far more autoregressive iterations than a typical medium-range weather forecast. Aurora is among a small number of AI weather models shown to remain stable over multi-year autoregressive rollouts in an independent benchmark spanning nine weather variables (Lehmann et al., 2026). We apply this noise separately to chemistry, radiation, and meteorological variables. Chemical concentrations receive a small positive lower bound of 10\textsuperscript{-30}. Radiative fluxes receive a lower bound of zero as radiative fluxes are non-negative and SW fluxes fall to zero at night and during polar winter. Training requires approximately 110 GH200 GPU hours, and optimization settings are given in Text S2.

\subsection{Testing Short-Term Predictions and Long Autoregressive Rollout}

Before evaluating the model autoregressively, we assess its short-term accuracy on the held-out 2014 test year. Each single-step prediction uses two consecutive days of UKCA output as input context and predicts the following day directly without autoregressive feedback. This isolates the model's one day ahead skill from any error that accumulates over a longer rollout and establishes the baseline against which the rollout results below are measured.

We evaluate the model autoregressively at three increasingly long horizons: one year, five years, and fifty years. The one-year rollout is initialized from UKCA January 1\textsuperscript{st} 2014 initial conditions and uses prescribed emissions over the year. The five-year and fifty-year rollouts recycle the same single test year's prescribed inputs repeatedly, while composition, meteorology, and radiation remain continuously autoregressive throughout. This tests whether the model remains stable beyond the horizon it was evaluated against. We assess stability at this longest horizon from year-over-year drift in global mean and zonal mean predicted-to-UKCA ratios for ozone, OH, and CO across several diverse pressure levels.

We report root mean square error (RMSE), normalized mean bias (NMB), and the coefficient of determination (R\textsuperscript{2}) defined in Text S3. All metrics are weighted by the cosine of latitude and by layer pressure thickness, and levels above each species' prescribed ceiling (Section 2.3) are excluded.

The emissions perturbation and satellite observation experiments each require additional training and risk degrading previously well-learned species or altitude ranges (i.e., ``catastrophic forgetting''; Kirkpatrick et al., 2017). Therefore, we branch each experiment from an earlier checkpoint in the training lineage rather than continue training from this fully converged model (Figure 1).

\subsection{Emissions Perturbation Experiments}

We investigate whether the model has learned a physically grounded response to emissions perturbations, rather than reproducing a fixed climatology regardless of forcing. Chemistry's response to an emissions change is nonlinear and regime dependent. For example, NO\textsubscript{x} both drives ozone production and can suppress ozone through titration depending on the background chemical regime. We use the NO-, CO-, and BC-scaled UKCA simulations described in Table 1.

Starting from an earlier checkpoint described in Section 2.5, we fine-tune on the perturbed emissions simulations to produce Aurora\textsubscript{UKCA,emiss}. We combine multiple emissions scenarios into a single training set and fine-tune for a single epoch at a learning rate of 1e-5. A single model therefore learns the emissions field as a continuous input forcing, allowing evaluation at untrained emission scalings and precursor combinations using one set of weights. We additionally introduce a paired-scenario loss that trains the model on its response to an emissions change directly, rather than only on the absolute state. We pair samples from the same calendar date in different UKCA emission scenarios and compute:

\begin{equation}
L_{\mathrm{sens}} = \frac{1}{K}\sum_{v=1}^{K}\frac{\left|\Delta_{\mathrm{pred}}(v) - \Delta_{\mathrm{target}}(v)\right|}{\sigma(v,l)}
\end{equation}

where Δpred and Δtarget are the predicted and target differences between the two scenarios' states, \emph{K} is the number of species evaluated, and σ is the root mean square tendency from Eq. 2. This term is added to the standard per-scenario loss for both members of the pair, including the emissions prediction loss described in Section 2.3. The emissions prediction loss supplies an additional gradient signal to the shared encoder and backbone, while the paired-scenario loss directly constrains differences in atmospheric predictions between emission scenarios.

We evaluate one-year rollouts driven by scaled daily emissions from the UKCA control simulation, including held-out years of scenarios used in training and scenarios withheld entirely from training (Table 1). During these rollouts, winds, temperature, geopotential height, and specific humidity are prescribed from the 2014 UKCA control simulation, while composition, aerosols, and radiation evolve autoregressively and predicted emissions are replaced with prescribed fields. Because the Aurora\textsubscript{UKCA,emiss} rollouts prescribe meteorology from the control simulation, their responses would differ slightly from UKCA's even if Aurora\textsubscript{UKCA,emiss} reproduced UKCA perfectly. This should be a second-order effect given the size of the emission scalings. Extending Aurora\textsubscript{UKCA,emiss} to fully free running simulations is left for future work.

\subsection{Satellite Ozone Observation Experiments}

We next test whether the same framework can learn from satellite observations by adding observed ozone profiles to the training objective, while retaining the physical, chemical, and climate relationships represented in UKCA.

We pair Aura MLS Level 2 ozone profile retrievals with a UKCA simulation whose winds and temperature are nudged toward ERA5 reanalysis. The MLS product spans 261 to 0.1 hPa, but we restrict our observational constraint to 250-20 hPa. Each retrieval returns a full vertical profile, such that each observed column is sampled at every level and the retrievals cover approximately 1.3\% of the horizontal domain. As such, ozone observations are sparse horizontally and dense vertically, which affects the loss formulation below. We train on nudged UKCA output and collocated MLS retrievals from 2012-2013 and hold out 2009-2011 for testing. Nudging aligns UKCA with the observed synoptic evolution, allowing simulated states to be associated with MLS retrievals from the same dates. A free-running simulation has no such correspondence, because its day-to-day weather is internally generated and cannot be matched to a specific date's satellite overpass.

We incorporate MLS observations through a sparse auxiliary loss (L\textsubscript{obs}) added to the standard UKCA training objective. For each model pressure level within this range, we select the nearest MLS retrieval level by pressure and compare the two without vertical interpolation. Before comparison, we convert predicted ozone from mass mixing ratio to volume mixing ratio using the molar masses of air and ozone, 28.96 and 48.00 g/mol, respectively. The observation loss is then:

\begin{equation}
L_{\mathrm{obs}} = \frac{1}{L}\sum_{l=1}^{L}\frac{1}{\kappa N_{l}}\sum_{i\in\Omega_{l}}\left|\mathrm{VMR}_{\mathrm{pred},i,l} - \mathrm{VMR}_{\mathrm{obs},i,l}\right|
\end{equation}

where Ω\textsubscript{l} is the set of grid cells containing a valid retrieval at level \emph{l} on that day, N\textsubscript{l} is its size, \emph{L} is the number of constrained levels observed that day, and κ is set to 5×10⁻⁶ mol/mol, which we find is a representative stratospheric volume mixing ratio scale. Grid-cell, level, and day combinations without MLS coverage are excluded from both sums. Averaging within a level before averaging across levels gives every observed level equal weight in the average, although each level's contribution still scales with its absolute ozone discrepancy because κ does not vary with altitude. This auxiliary loss is applied only to ozone. All other chemical species and meteorological variables continue to train against UKCA targets as described in Section 2.3.

To isolate the effect of observation-constrained training, we compare the MLS-informed model (Aurora\textsubscript{obs}) with a matched control (Aurora\textsubscript{control}) trained on the same nudged UKCA dataset using identical settings but with the MLS loss weight set to zero. The difference between these models measures the change attributable specifically to the MLS auxiliary loss. We refer to this as the MLS-induced model correction (rather than a conventional assimilation increment) because it results from changes to learned model parameters rather than the ingestion of concurrent observations during inference.

\section{Results and Discussion}

\subsection{Short-term prediction accuracy}

Aurora\textsubscript{UKCA} reproduces most of the coupled chemistry-climate state with high single-step accuracy over the test year (Tables S1-S2). Of the 27 evaluated chemical, meteorological, and radiative fields, 23 achieve an R\textsuperscript{2} of at least 0.85 in both the troposphere and stratosphere, and 19 achieve at least 0.90. All five meteorological fields (zonal and meridional winds, temperature, geopotential height, and specific humidity) and all four radiative fluxes exceed 0.90 in both regions. NMB remains within ±5\% for 17 of the 22 chemical and radiative fields. Tropospheric ozone achieves an R\textsuperscript{2} of 0.98, an NMB of +0.75\%, and an RMSE of 5.77 ppb, while stratospheric ozone exceeds an R\textsuperscript{2} of 0.99. Longer chemical lifetimes and slower transport in the lower stratosphere likely aid daily prediction by preserving concentration patterns between time steps.

Radiative accuracy is strongest for downwelling fluxes with relatively larger errors in reflected SW radiation. Tropospheric downwelling LW and SW radiation achieve R\textsuperscript{2} values of 0.99 and 0.98, with RMSEs of 8.58 and 16.8 W m⁻², respectively. Upwelling SW radiation reaches R\textsuperscript{2} values of 0.92 in the troposphere and 0.91 in the stratosphere. Incoming sunlight strongly constrains downwelling radiation near the domain top, whereas upwelling radiation depends on reflection by clouds, aerosols, and the surface below. The model places the low-cloud variable correctly in 95.5\% of grid cells and predicts a cloud extent of 18.0\% against UKCA's 17.9\% (Figure S1). Because this binary variable cannot contain the gradations in cloud fraction that determine albedo, reducing cloud to a thresholded field limits how well reflected SW can be reproduced even where this variable is correct.

Annual global mean vertical profiles generally agree with UKCA to within ±10\% (Figures S2-5). Profiles of the oxidants (ozone, OH, HO\textsubscript{2}) and CO remain within this range at every predicted level as do all four radiative fluxes. Larger departures occur mainly in the upper atmosphere where several species decline steeply with altitude. Cl\_precursor concentrations are reproduced to better than 2\% through the troposphere and lower stratosphere but decrease by three orders of magnitude above 30 hPa through photolysis. Its 25\% underestimate at 1 hPa occurs where concentrations have fallen to 0.1\% of the column maximum. This pattern suggests that the model captures the relatively uniform Cl\_precursor reservoir more accurately than its transition to photochemical depletion. In this steeply declining part of the profile, even a small vertical displacement could produce a large relative concentration error. SO\textsubscript{2} is a broader exception with errors exceeding 10\% at five of the 20 levels. Its largest underestimate of 37\% occurs at 100 hPa, near the sharp minimum between its tropospheric and stratospheric reservoirs.

Aurora\textsubscript{UKCA} captures the vertical and latitudinal structure of chemically active species (Figure 2). Across the annual mean ozone cross-section, the median relative difference from UKCA is +0.79\% with 90\% of differences within ±5.0\%. ClOₓ agrees with UKCA to within approximately ±10\% throughout the stratosphere where concentrations are largest. At 925 hPa, CO achieves an R\textsuperscript{2} of 0.98, an NMB of +1.59\%, and an RMSE of 4.37 ppb. NOₓ achieves an R\textsuperscript{2} of 0.94, an NMB of -1.29\%, and an RMSE of 0.14 ppb. Its largest relative errors occur at low concentrations over oceans and remote regions, resembling difficulties reported for other learned chemical models (Keller and Evans, 2019; Kelp et al., 2022; Xia et al., 2025). Despite these localized NOₓ errors, ozone and CO retain high prediction accuracy. The maps also exhibit patch artifacts aligned with the encoder's grid. This pattern reflects the model's reconstruction of continuous atmospheric fields from discrete patches (Han et al., 2023). These patch artifacts are more pronounced for species with short correlation length scales such as NOₓ compared to longer ones such as ozone.

\begin{figure}[htbp]
\centering
\includegraphics[width=\textwidth,height=0.82\textheight,keepaspectratio]{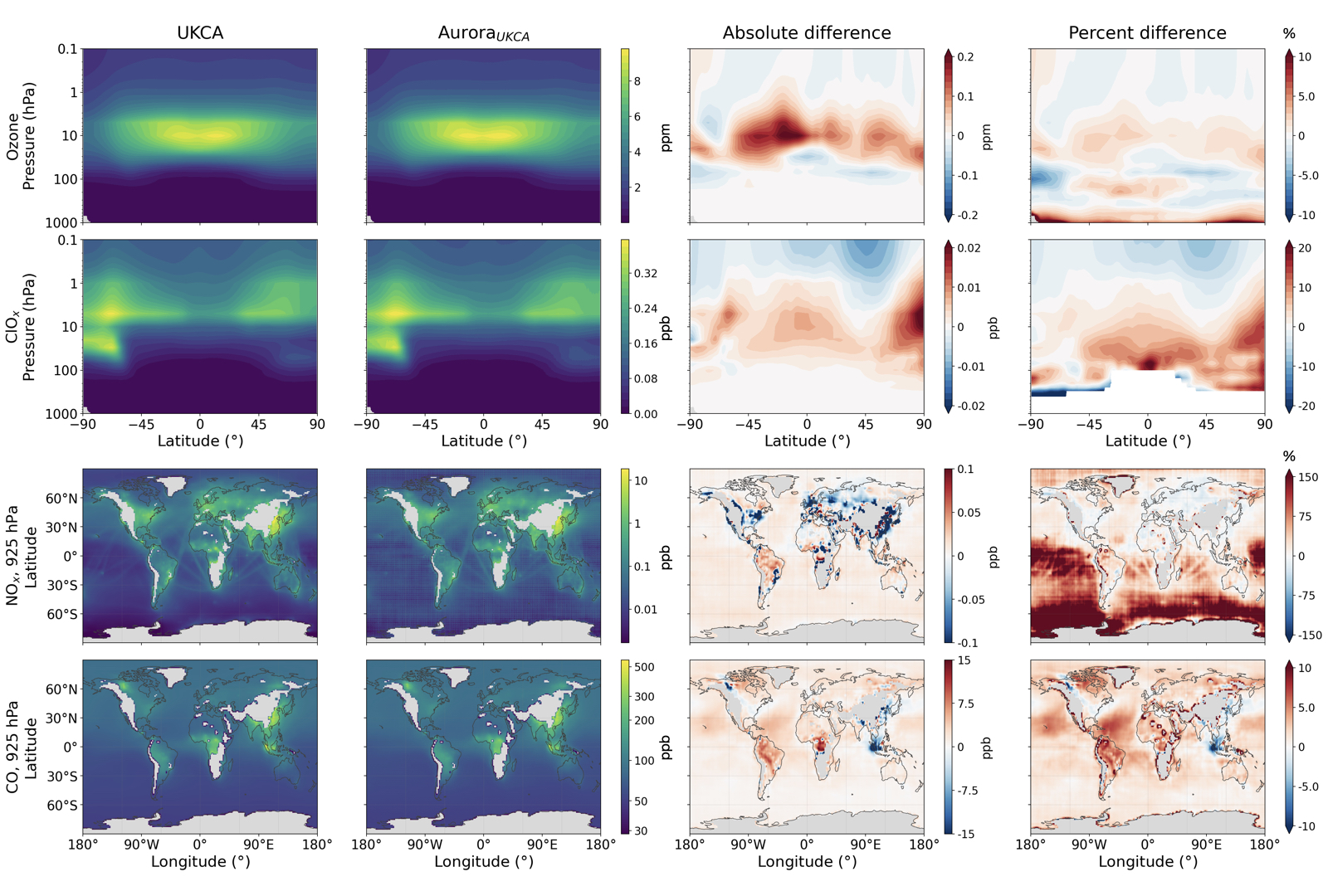}
\caption{\textbf{Annual mean atmospheric composition}. Columns show the UKCA reference, Aurora\textsubscript{UKCA} prediction, and the absolute and relative differences. The top two rows show zonal mean cross-sections of ozone and ClO\textsubscript{x}, and the bottom two rows show NO\textsubscript{x} and CO at 925 hPa. Fields are averaged over the 358 single-step evaluation days in 2014. Gray cells indicate locations where the pressure level lies below the model surface and UKCA reports zero concentration. White regions in the ClOₓ panels indicate concentrations excluded from the training loss and masked in the figure.}
\label{fig:2}
\end{figure}

In Aurora\textsubscript{UKCA}, SO\textsubscript{2} is reproduced less accurately than other chemical species. Its R\textsuperscript{2} is 0.68 in the troposphere and 0.70 in the stratosphere with NMBs of -1.78\% and -10.7\%, respectively. One possible explanation is that our selected variables provide incomplete information about SO\textsubscript{2} sources and sinks. SO\textsubscript{2} emissions are not supplied as inputs, and neither its precursors (e.g., DMS, COS) nor its oxidation product H\textsubscript{2}SO\textsubscript{4} is predicted. Sulfate aerosol is predicted only in the soluble accumulation mode. Aurora\textsubscript{UKCA} must therefore infer the effects of these omitted sources and sinks from the retained SO\textsubscript{2} and sulfate fields alone. Kelp et al. (2022) also found poorer H\textsubscript{2}SO\textsubscript{4} predictions in a simplified mechanism that lacked chemical loss for that species, allowing prediction biases to accumulate. That limitation arose from the chemical budget represented in the reference mechanism. Here, SO\textsubscript{2} sources and sinks are represented in UKCA, but the selected variables may provide insufficient information for Aurora\textsubscript{UKCA} to learn their combined effects.

\subsection{Long-term stability and variability}

Aurora\textsubscript{UKCA} preserves the seasonal cycle over five years with stable predictions and small changes in its mean bias (Figure 3). Averaged across a five-year horizon, ozone, CO, and OH at 500 hPa are biased low by 3.0\%, 2.0\%, and 0.9\%, respectively, while outgoing LW radiation is low by 0.6\%. Between the first and fifth years, these biases change by only +0.16\%, -0.05\%, -0.30\%, and -0.05\%. The corresponding seasonal bias cycles for all predicted chemical and radiative fields remain similarly bounded with annual means within ±7\% for every field (Figure S6). The global mean vertical structure of every predicted field also remains consistent across the five annual cycles (Figure S7-8). Nevertheless, successive years display variability despite the underlying repeated annual forcing. This variability corresponds to 10-25\% of the standard deviation of the respective seasonal cycles. Daily variability is somewhat weaker than in UKCA with the standard deviation of daily global mean changes reaching 78-88\% of UKCA's value for ozone, CO, and outgoing LW radiation. We find that the model maintains a stable seasonal mean while producing year-to-year differences under the same repeated annual forcing.

\begin{figure}[htbp]
\centering
\includegraphics[width=\textwidth,height=0.82\textheight,keepaspectratio]{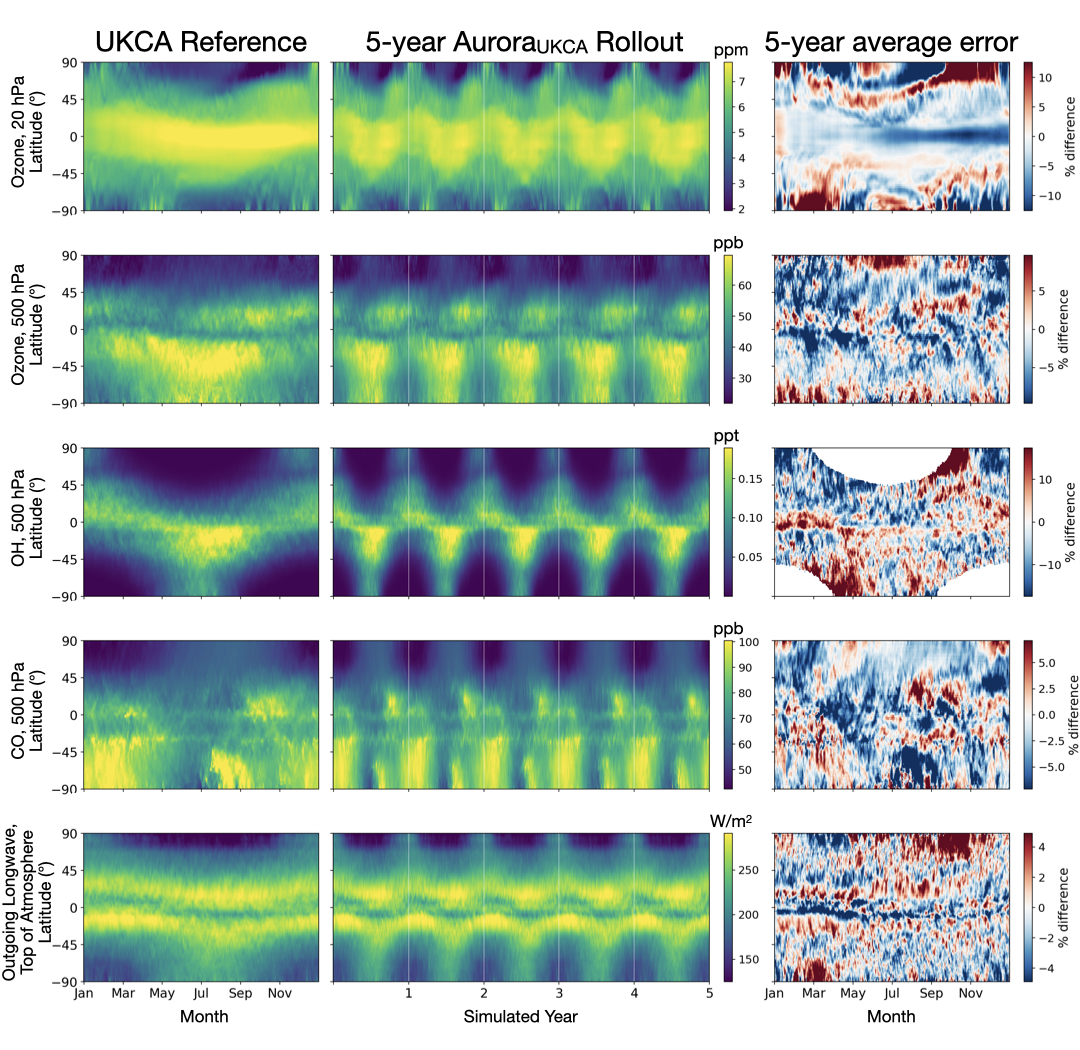}
\caption{\textbf{Autoregressive rollouts of the fine-tuned Aurora model compared with UKCA}. Zonal mean Hovmöller diagrams for ozone, OH, CO, and outgoing LW radiation at the top of the atmosphere. The left column shows the UKCA daily target over the 12 months of the 2014 test year. The middle column shows the five-year free running evolution of the rollout given only initial conditions. The right column shows the mean percent difference averaged over the five annual cycles. The 2014 test year is repeated as forcing, so a model free of drift reproduces the same annual cycle five times. Only boundary conditions (surface meteorology and emissions) are prescribed. Percent differences are masked where the UKCA target falls below 5\% of its 98th percentile, which removes polar night for OH.}
\label{fig:3}
\end{figure}

Accuracy varies across regions during rollouts. Polar mean biases reach up to +9.5\% for OH, and +7.7\% for downwelling SW radiation. Ozone is less affected with a polar bias of -2.4\% compared with -0.4\% in the tropics and +0.3\% in the midlatitudes. Reduced polar skill has also been reported for GraphCast, which underperforms ECMWF HRES near the poles for specific humidity, geopotential, and 2-m temperature (Lam et al., 2023). In Aurora\textsubscript{UKCA}, strong seasonal contrasts between winter darkness and summer sunlight may also make polar radiation and photochemistry harder to reproduce consistently. The largest relative OH errors occur during polar night when UKCA concentrations approach zero and even small absolute differences can produce large relative errors.

The 50-year rollout also maintains stable seasonal cycles (Figures S9-13). Trends in annual NMB remain within ±0.09 percentage points per decade for all variables in Figure 3. The annual NMBs fluctuate between cycles with standard deviations of 0.04 to 0.41 percentage points but show no statistically significant long-term trend at the 95\% confidence level, consistent with the internal variability described above as opposed to progressive drift. These results suggest that single-step fine-tuning with input noise can support simulations lasting decades. A one-year rollout of 358 daily steps completes in approximately 5.5 minutes on a single NVIDIA GH200 GPU. The equivalent UKCA AMIP integration with Strat-Trop chemistry requires approximately 9.5 hours on 576 CPU cores (Monsoon3). Therefore, the time to solution is reduced by a factor of around 100 with arguably much less compute hardware. These are wall clock comparisons on different hardware rather than a normalized measure of computational work.

Aurora\textsubscript{UKCA} avoids explicit operator splitting during inference by jointly predicting the daily UKCA outputs. Conventional CCMs evolve atmospheric processes through separate numerical operators with each operator acting on the state produced by preceding processes. Replacing the chemical solver with an AI-learned operator retains this coupling and requires the learned component to remain accurate under conditions repeatedly modified by transport and physics. Earlier studies reported error accumulation during repeated chemical integration (Kelp et al., 2018) and within numerical transport models (Keller and Evans, 2019). Aurora\textsubscript{UKCA}'s training targets reflect the combined daily effects of these processes, allowing their evolution to be learned within a single model system. Removing the interface between learned chemistry and numerical transport may reduce inconsistencies introduced by their coupling and contribute to the observed stability. Prediction errors can nevertheless accumulate through autoregressive feedback. We find that input noise during fine-tuning may help limit this risk by exposing the model to perturbations around the UKCA training states, reducing the mismatch between the states encountered during training and those generated during rollout. Although avoiding explicit operator splitting may support stability, jointly representing chemical and physical processes makes their individual contributions harder to identify. This opacity complicates interpretation and the diagnosis of prediction errors (Hu et al., 2026), while also raising concerns regarding AI model governance and trust (Kelp et al., 2026).

\subsection{\texorpdfstring{Fine-tuning for NO\textsubscript{x}-O\textsubscript{x}-VOC Chemistry}{Fine-tuning for NOx-Ox-VOC Chemistry}}

Aurora\textsubscript{UKCA,emiss} reproduces the regional ozone response to NO emission scenarios withheld from training (Figure 4). For the 1.5×NO emission scenario, it captures widespread ozone increases at 925 hPa and the decrease over eastern China, where background NO emissions are high. The trained 2× and 0.5×NO emission scenarios show the same spatial contrast (Figure S14). The model has thus learned an ozone sensitivity to NO that changes sign by region rather than a uniform response. This contrast is consistent with the model distinguishing between chemical regimes, although annual mean ozone itself does not separate NO titration from changes in net photochemical production. However, the positive response is too strong over parts of tropical Africa and South America. These regions combine high biogenic VOC emissions with low background NO\textsubscript{x}, where ozone production is most sensitive to added NO\textsubscript{x}. The overestimate may therefore reflect an exaggerated sensitivity in this NO\textsubscript{x}-limited regime.

Ozone responses to CO are less accurate. For the withheld 1.5×CO emission scenario, Aurora\textsubscript{UKCA,emiss} generally simulates the expected increase in ozone but this increase is biased high relative to UKCA. Increases are simulated over East Asia and India in Aurora\textsubscript{UKCA,emiss} but spatially these do not align exactly with the increases in UKCA. The trained scenarios show similar disagreement (Figure S14). One possible explanation is CO's lifetime of one to two months, which spreads its influence on ozone far from emission sources. Each daily step then contains only a small CO-driven ozone increment relative to day-to-day variability, providing a weak signal under our tendency-normalized loss (Section 2.3). How emissions are considered during training also affects these responses. We find that supplying emissions only as inputs produces the least accurate responses, supervising them as outputs improves them, and the paired-scenario loss provides a further improvement by directly constraining differences between scenarios. The paired-scenario loss likely helps because it isolates the emission-driven change from shared day-to-day variability. Responses to long-lived precursors such as CO would therefore benefit most from longer perturbed simulations.

Aurora\textsubscript{UKCA,emiss} can reproduce the ozone response to a combination of emission changes absent from its training. Every training scenario perturbs a single emissions precursor. Yet for the combined doubling of NO and CO, the model reproduces the smaller region of ozone decrease over eastern China that UKCA simulates relative to the NO-only perturbation (Figure 4). This behavior is consistent with CO oxidation supplying HO\textsubscript{2}, which converts NO to NO\textsubscript{2} without consuming ozone. Reproducing it requires the model to combine sensitivities learned separately for each precursor and it does so even though its response to CO alone is less accurate. Zonal mean responses show that agreement does not extend consistently through the column (Figures S15-S16). The model captures broad ozone increases through the lower and middle troposphere, including the stronger Northern Hemisphere response to combined increases. However, it disagrees near 200-300 hPa with opposite signs at southern high latitudes in the individual 1.5×NO and CO emission experiments. Near the tropopause, steep vertical ozone gradients allow small displacements to produce large differences and the signal far from emission sources is weak.

\begin{figure}[htbp]
\centering
\includegraphics[width=\textwidth,height=0.82\textheight,keepaspectratio]{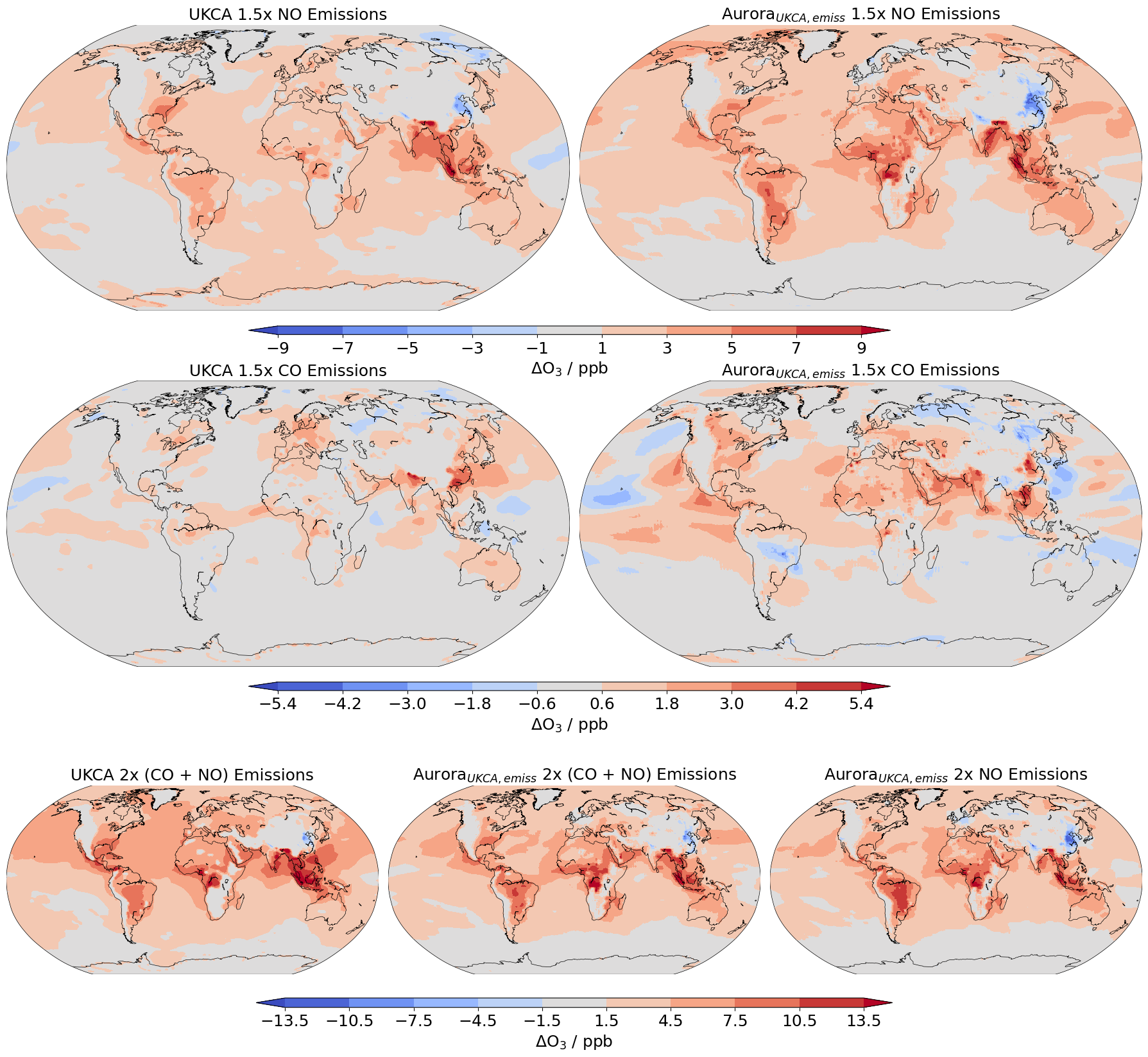}
\caption{\textbf{Ozone responses to emission scenarios withheld from training.} Change in annual mean ozone mole fraction at 925 hPa relative to the control following scalings of 1.5 to NO emissions (top row), 1.5 to CO emissions (middle row), 2 to CO and NO emissions and 2 to NO emissions (bottom row).}
\label{fig:4}
\end{figure}

\subsection{Fine-tuning for Aerosol-Radiation Interactions}

Aurora\textsubscript{UKCA,emiss} reproduces the increase in BC concentration following a tripling of BC emissions (Figure 5). This scaling lies between the 2× and 5× scalings used in training, so the model must interpolate its learned response to an emission level unseen. In both models, the largest increases occur over biomass burning regions such as central Africa and industrial regions such as China. BC's short atmospheric lifetime keeps its response close to these emission sources, which may make it easier to learn than the more widely distributed ozone response to CO (Section 3.3). Because BC absorbs SW radiation and re-emits LW radiation, these increases reduce the SW radiation reaching the surface and alter the atmosphere's radiative balance. Aurora\textsubscript{UKCA,emiss} captures this reduction over the same regions, predominantly Africa and East Asia, but underestimates its magnitude relative to UKCA.

The underestimated radiative response likely reflects the short duration of the UKCA training simulations. The full radiative response to a BC perturbation includes both the instantaneous radiative effect and the subsequent rapid adjustments in temperature, humidity, and clouds. With only two years of perturbed training simulations, these adjustments are difficult to separate from meteorological variability and therefore difficult to learn. Prescribing meteorology from the control simulation during these rollouts may also contribute, as Aurora\textsubscript{UKCA,emiss} cannot represent the temperature and humidity adjustments that accompany the BC increase in UKCA. As such, Aurora\textsubscript{UKCA,emiss} captures the spatial pattern of the surface SW response but we do not calculate the effective radiative forcing (ERF).

\begin{figure}[htbp]
\centering
\includegraphics[width=\textwidth,height=0.82\textheight,keepaspectratio]{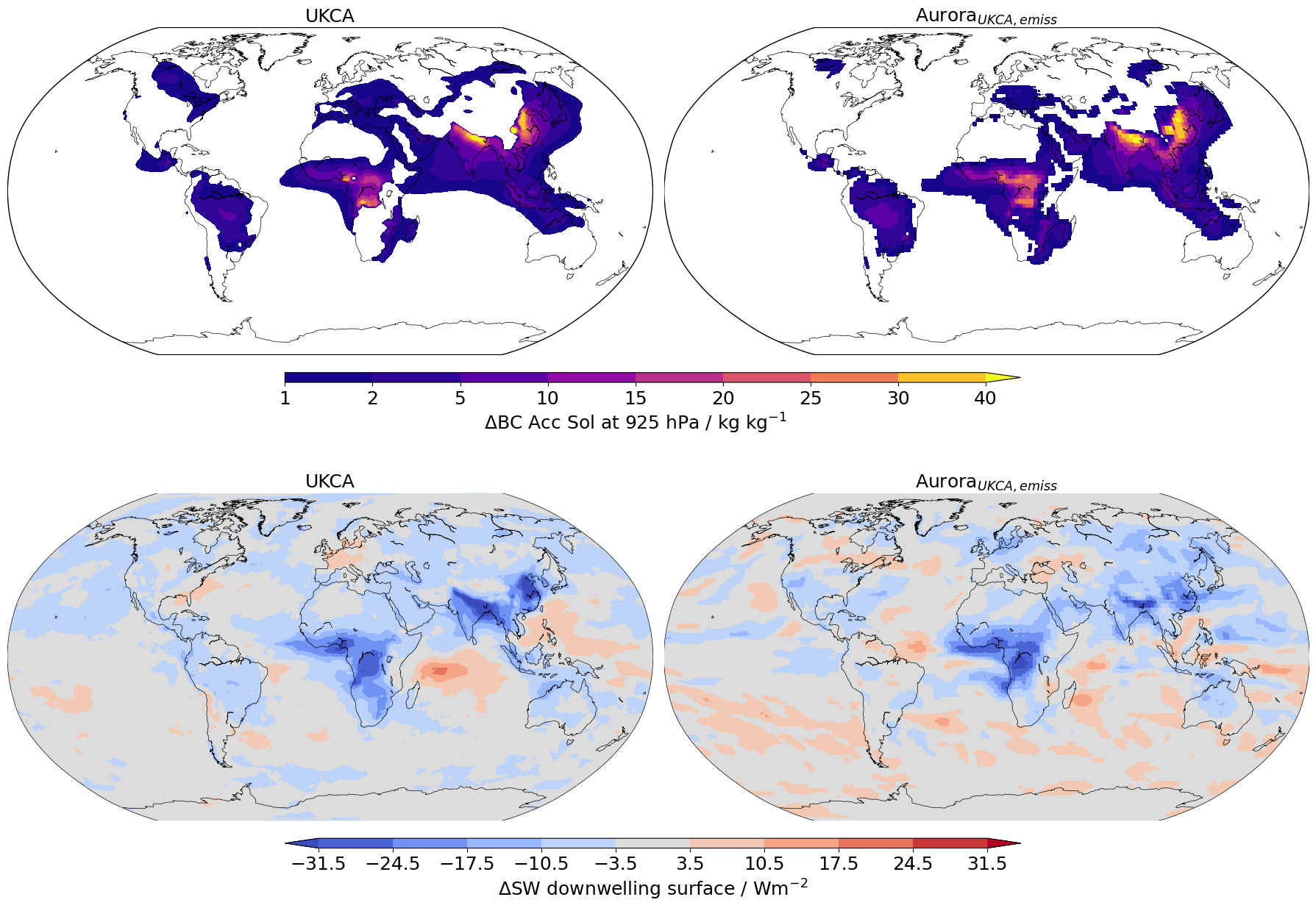}
\caption{\textbf{Response of aerosol and radiation to BC emission changes.} Annual mean change relative to the control following a withheld 3× scaling of BC emissions. BC mass mixing ratio in the soluble accumulation mode at 925 hPa (top row) and surface downwelling SW radiation (bottom row), for UKCA (averaged over 2014-2017) and Aurora\textsubscript{UKCA,emiss}.}
\label{fig:5}
\end{figure}

\subsection{Fine-tuning on Satellite Ozone Retrievals}

Observation-constrained training improves the stratospheric zonal mean ozone distribution across the test period (Figure 6). Calculated against collocated MLS retrievals binned by latitude and pressure level between 250 and 20 hPa, RMSE decreases by 23.4\% from 0.37 ppm in Aurora\textsubscript{control} to 0.28 ppm in Aurora\textsubscript{obs}. Improvements occur in all four seasons, with larger reductions in JJA (32.0\%) and DJF (25.3\%) than in MAM (18.6\%) and SON (18.5\%) (Figures S17-S19). In JJA, RMSE decreases from 0.38 to 0.26 ppm with improvements in 70\% of latitude bins. The reduction reaches 49.2\% in the Northern Hemisphere against 16.5\% in the Southern Hemisphere. The learned correction follows the broad meridional and vertical pattern of the initial MLS model discrepancy (Figure 6B and 6C), bringing the zonal mean ozone distribution closer to the observations. We observe that the observational constraint is most effective when the stratosphere is dynamically calm and less effective during dynamically active polar spring conditions.

\begin{figure}[htbp]
\centering
\includegraphics[width=\textwidth,height=0.82\textheight,keepaspectratio]{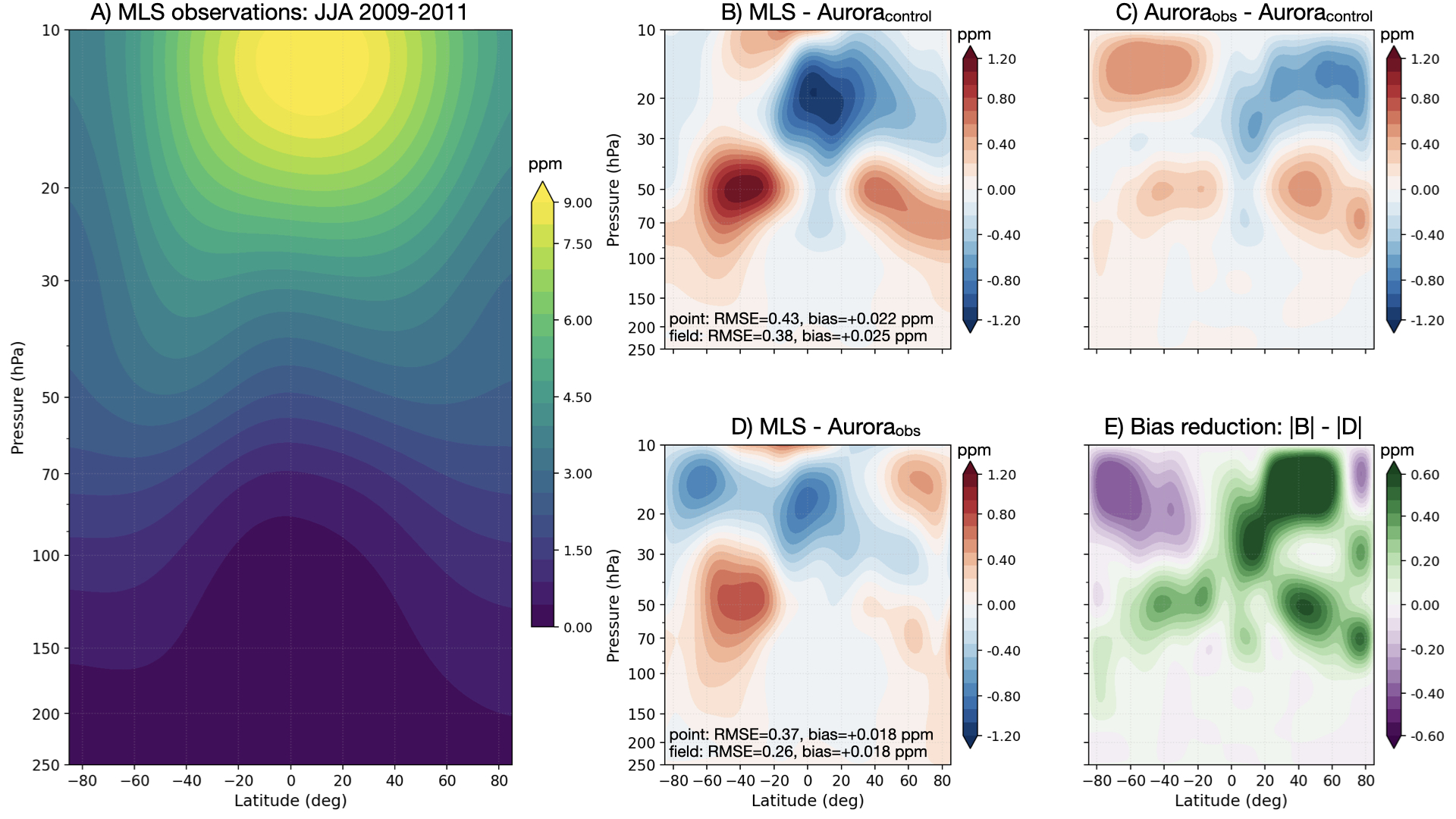}
\caption{\textbf{Zonal mean ozone for June through August averaged over the held-out 2009-2011 period.} (A) Ozone observed by MLS. (B) MLS minus the nudged Aurora simulation (Aurora\textsubscript{control}), representing the discrepancy before the observational constraint is applied. (C) Aurora\textsubscript{obs} minus Aurora\textsubscript{control}, the correction learned from the observations. (D) MLS minus Aurora\textsubscript{obs}, representing the remaining discrepancy. (E) \textbar B\textbar{} − \textbar D\textbar, green where the model moves toward MLS and purple where it moves away. All quantities are in ppm, and positive values in (B) and (D) indicate that MLS exceeds the model.}
\label{fig:6}
\end{figure}

The improvement in zonal mean ozone is accompanied by degraded performance at individual collocated retrievals. Aurora\textsubscript{obs} has a 9.6\% higher RMSE than Aurora\textsubscript{control} across 29 million matched points (0.47 vs 0.43 ppm). Mean biases remain similar at approximately +0.02 ppm, indicating that the higher RMSE reflects a greater spread of errors across individual collocations. Zonal averaging can reduce errors of opposite sign, allowing the mean distribution to improve even when errors at individual retrievals increase. JJA is the only season in which both metrics improve with 32.0\% for the zonal mean and 14.4\% at individual retrievals. Retrieval-level RMSE increases in the other three seasons, including increases of 18.6\% in DJF and 24.7\% in SON despite improved zonal means. Therefore, this observation-constrained fine-tuning method is suited to applications that depend on the zonal mean or climatological ozone distribution rather than on instantaneous, local ozone profiles. Because the correction is encoded in the learned weights and requires no observations at inference time, it transfers to simulations for which no retrievals exist. This property distinguishes observational fine-tuning from conventional data assimilation.

The learned correction does not transfer beyond the levels where the observations are applied, despite the shared model weights used to predict ozone throughout the domain. In theory, because L\textsubscript{obs} propagates through the entire model architecture (i.e., encoder, backbone, and decoder), an observational constraint imposed between 250 and 20 hPa could influence ozone at all levels. The model may satisfy this constraint either by generalizing the correction upward or by redistributing ozone through the column. However, we observe neither response. Above 20 hPa, Aurora\textsubscript{obs} remains within 6\% of Aurora\textsubscript{control}. Furthermore, the adjustment is uneven within the constrained range. The zonal mean error decreases by 19 to 28\% between 20 and 70 hPa, where ozone concentrations reach several ppm and Aurora\textsubscript{control}'s error is largest. At 100 and 200 hPa, where ozone concentrations fall below approximately 1 ppm, error increases by 9.5\% and 20\%, respectively. However, the corresponding absolute errors remain relatively small at 0.03 to 0.20 ppm across the two models. This vertical contrast is consistent with the constant normalization in Eq. 7. With κ fixed across levels, larger absolute ozone discrepancies contribute more to the observation loss despite equal nominal weights for each level. The improvement in the overall zonal mean metric may thus conceal weaker performance in the lower concentration parts of the profile.

\section{Conclusions}

We present the first foundation model adapted for interactive chemistry-climate simulation. We fine-tune the Aurora Earth system foundation model on UKCA output and evaluate its accuracy, stability, emission responses, and ability to learn from satellite observations. This combined analysis shows that a pretrained weather foundation model can be converted into a stable CCM primarily through its training objective. Loss functions become the primary avenue through which we adapt the pretrained model, determining how species with widely differing lifetimes are weighted and how emissions are jointly optimized with composition, meteorology, and radiation. These losses guide what the model learns from UKCA rather than imposing governing chemistry, so we ultimately evaluate physical consistency after training. The resulting model reproduces most fields with high accuracy while running about 100× faster than UKCA. Joint prediction avoids explicit operator splitting and single step training with input noise supports stability over 50-year rollouts, whereas optimizing over specific time windows is less stable and ties the model to the training rollout length. Aurora\textsubscript{UKCA,emiss} reproduces the spatial response of ozone to NO\textsubscript{x} in both NO\textsubscript{x}-limited and VOC-limited regions and of surface SW radiation to BC. Finally, we demonstrate the first known use of chemical observations to fine-tune a foundation model for chemistry-climate simulation. Training on MLS ozone retrievals corrects inherited UKCA biases in the zonal mean ozone distribution. Because this correction is encoded in the learned weights, it can transfer to simulations for which no observations exist.

Most limitations of Aurora\textsubscript{UKCA} trace back to the scope and availability of its UKCA training data. We deliberately predict a subset of UKCA's more than 100 coupled gas and aerosol species to keep training within hardware memory and storage limits. However, omitted variables limit what the model can infer about the species it predicts. SO\textsubscript{2} is the least accurate species, likely because its emissions, precursors, and principal chemical sinks are not included. The short duration of the perturbed emission simulations also limits how well Aurora\textsubscript{UKCA,emiss} learns slower-evolving responses. Its ozone response to CO is less accurate than its response to NO\textsubscript{x}, possibly because of CO's longer lifetime, and it underestimates the surface radiative response to BC. Because the 50-year rollouts recycle a single year of forcing, their stability under evolving emissions and climate remains untested.

Addressing these limitations depends less on AI model architecture than on training data that is costly to generate. Every additional species, scenario, or simulated year requires more UKCA simulation and storage. As such, the speedup of Aurora\textsubscript{UKCA} is realized only after a large upfront investment in process-based simulation. Learning the rapid adjustments that determine ERF is a natural next step, but it will require perturbed emission simulations far longer than those used here. Because ERF is critical to climate policy assessment, this capability would allow Aurora\textsubscript{UKCA} to span emission scenarios that are too costly to simulate with process-based models alone. Beyond ERF, diagnosing the full climate response to a perturbation will require coupling Aurora\textsubscript{UKCA}, which currently behaves as an atmosphere-only model, to physical or AI models of the rest of the Earth system, analogous to the coupling of UKCA within UKESM. Observational fine-tuning offers a complementary path to model improvement that does not require new UKCA simulations. However, the sparse coverage of chemical observations limits how far this approach can extend beyond well-observed species such as ozone (Kelp et al., 2025).

Further development of Aurora\textsubscript{UKCA} also requires evaluation against experiments that define CCM performance. AerChemMIP, AerChemMIP2, and RAMIP (Collins et al., 2017; Fiedler et al., 2026; Wilcox et al., 2023) provide standard benchmarks, including single species perturbations (e.g., piClim-NO\textsubscript{x} vs. piClim-control) and leave-one-species-out transient simulations (e.g., histSST relative to histSST-piCH\textsubscript{4}). Their fixed SST experiments match AuroraUKCA's atmosphere-only configuration and the piClim experiments are the standard framework for diagnosing ERF. Therefore, reproducing the pivotal responses simulated across models in these MIPs is a prerequisite for using Aurora\textsubscript{UKCA} to investigate chemistry-aerosol-climate interactions. As other AI CCMs emerge from different combinations of foundation models and CCM training data, a dedicated intercomparison could extend AIMIP (Henn et al., 2026) beyond the physical climate to chemistry-climate and air quality responses (Griffiths et al., 2025). The low cost of AI models would allow such an intercomparison to span a far wider parameter space than AerChemMIP2 (Hickman et al., 2025).

Aurora\textsubscript{UKCA} does not replace process-based models such as UKCA, nor does it obviate their development. Because AI models trained on simulation output inherit both the physical consistency and the missing processes of their parent models, more faithful representations of chemical, aerosol, and radiative processes yield higher quality training data and more skillful AI models (Kelp et al., 2026). Training on observations alongside simulations, as in our MLS experiments, may in turn reveal gaps in the parent model. This relationship also depends on a shared standard for reporting the physical behavior of AI models in Earth science. Agreement with reference fields alone does not establish that a model represents the processes it predicts (Hu et al., 2026). Because jointly predicting the full state makes individual chemical and physical processes harder to identify, we cannot determine from these experiments whether Aurora\textsubscript{UKCA}'s emission responses reflect chemical mechanisms or statistical regularities in its training data. Pairing mechanistic interpretability with UKCA process diagnostics could test this by linking the model's internal features to known atmospheric processes. Such a standard would document both physical tests like those presented here and analyses of internal mechanisms and treat preprocessing choices as part of the model.

\medskip
\textbf{Computing resources:} The model training and rollouts were conducted on the Jasmin Orchid cluster and Isambard-AI HPC.

\textbf{Competing interests:} The authors declare there are no conflicts of interest for this manuscript.

\textbf{Data and materials availability:} The model checkpoints and data used to make the plots are available on the Zenodo repository \href{https://doi.org/10.5281/zenodo.22855213}{https://doi.org/10.5281/zenodo.22855213}.

\clearpage
\section*{References}
\begin{list}{}{\leftmargin=1.5em\itemindent=-1.5em\itemsep=2pt\parsep=0pt}\small
\item Allen, R.J., Horowitz, L.W., Naik, V., Oshima, N., O'Connor, F.M., Turnock, S., Shim, S., Le Sager, P., van Noije, T., Tsigaridis, K., Bauer, S.E., Sentman, L.T., John, J.G., Broderick, C., Deushi, M., Folberth, G.A., Fujimori, S., Collins, W.J., 2021. Significant climate benefits from near-term climate forcer mitigation in spite of aerosol reductions. Environ. Res. Lett. 16, 034010. \url{https://doi.org/10.1088/1748-9326/abe06b}
\item Archibald, A.T., O'Connor, F.M., Abraham, N.L., Archer-Nicholls, S., Chipperfield, M.P., Dalvi, M., Folberth, G.A., Dennison, F., Dhomse, S.S., Griffiths, P.T., Hardacre, C., Hewitt, A.J., Hill, R.S., Johnson, C.E., Keeble, J., Köhler, M.O., Morgenstern, O., Mulcahy, J.P., Ordóñez, C., Pope, R.J., Rumbold, S.T., Russo, M.R., Savage, N.H., Sellar, A., Stringer, M., Turnock, S.T., Wild, O., Zeng, G., 2020. Description and evaluation of the UKCA stratosphere--troposphere chemistry scheme (StratTrop vn 1.0) implemented in UKESM1. Geoscientific Model Development 13, 1223--1266. \url{https://doi.org/10.5194/gmd-13-1223-2020}
\item Bi, K., Xie, L., Zhang, H., Chen, X., Gu, X., Tian, Q., 2023. Accurate medium-range global weather forecasting with 3D neural networks. Nature 619, 533--538. \url{https://doi.org/10.1038/s41586-023-06185-3}
\item Bocquet, M., Elbern, H., Eskes, H., Hirtl, M., Žabkar, R., Carmichael, G.R., Flemming, J., Inness, A., Pagowski, M., Pérez Camaño, J.L., Saide, P.E., San Jose, R., Sofiev, M., Vira, J., Baklanov, A., Carnevale, C., Grell, G., Seigneur, C., 2015. Data assimilation in atmospheric chemistry models: current status and future prospects for coupled chemistry meteorology models. Atmospheric Chemistry and Physics 15, 5325--5358. \url{https://doi.org/10.5194/acp-15-5325-2015}
\item Bodnar, C., Bruinsma, W.P., Lucic, A., Stanley, M., Allen, A., Brandstetter, J., Garvan, P., Riechert, M., Weyn, J.A., Dong, H., Gupta, J.K., Thambiratnam, K., Archibald, A.T., Wu, C.-C., Heider, E., Welling, M., Turner, R.E., Perdikaris, P., 2025. A foundation model for the Earth system. Nature 641, 1180--1187. \url{https://doi.org/10.1038/s41586-025-09005-y}
\item Brenowitz, N.D., Henn, B., McGibbon, J., Clark, S.K., Kwa, A., Perkins, W.A., Watt-Meyer, O., Bretherton, C.S., 2020. Machine Learning Climate Model Dynamics: Offline versus Online Performance. \url{https://doi.org/10.48550/arXiv.2011.03081}
\item Collins, W.J., Lamarque, J.-F., Schulz, M., Boucher, O., Eyring, V., Hegglin, M.I., Maycock, A., Myhre, G., Prather, M., Shindell, D., Smith, S.J., 2017. AerChemMIP: quantifying the effects of chemistry and aerosols in CMIP6. Geoscientific Model Development 10, 585--607. \url{https://doi.org/10.5194/gmd-10-585-2017}
\item Dewey, M., Hansson, H.-C., Watson-Parris, D., Samset, B.H., Wilcox, L.J., Lewinschal, A., Sand, M., Seland, Ø., Krishnan, S., Ekman, A.M.L., 2025. AeroGP: Machine Learning How Aerosols Impact Regional Climate. Journal of Geophysical Research: Machine Learning and Computation 2, e2025JH000741. \url{https://doi.org/10.1029/2025JH000741}
\item Eastham, S.D., Long, M.S., Keller, C.A., Lundgren, E., Yantosca, R.M., Zhuang, J., Li, C., Lee, C.J., Yannetti, M., Auer, B.M., Clune, T.L., Kouatchou, J., Putman, W.M., Thompson, M.A., Trayanov, A.L., Molod, A.M., Martin, R.V., Jacob, D.J., 2018. GEOS-Chem High Performance (GCHP v11-02c): a next-generation implementation of the GEOS-Chem chemical transport model for massively parallel applications. Geoscientific Model Development 11, 2941--2953. \url{https://doi.org/10.5194/gmd-11-2941-2018}
\item Fan, H., Bai, L., Fei, B., Xiao, Y., Chen, K., Liu, Y., Qu, Y., Ling, F., Gentine, P., 2026. Physically consistent global atmospheric data assimilation with machine learning in latent space. Science Advances 12, eaea4248. \url{https://doi.org/10.1126/sciadv.aea4248}
\item Fiedler, S., O'Connor, F.M., Watson-Parris, D., Allen, R.J., Collins, W.J., Griffiths, P.T., Kasoar, M., Kikstra, J., Kok, J.F., Murray, L.T., Paulot, F., Sand, M., Turnock, S.T., Weber, J., Wilcox, L.J., Naik, V., 2026. AerChemMIP2 -- unraveling the role of reactive gases, aerosol particles, and land use for air quality and climate change in CMIP7. Geoscientific Model Development 19, 3477--3508. \url{https://doi.org/10.5194/gmd-19-3477-2026}
\item Griffiths, P.T., Wilcox, L.J., Allen, R.J., Naik, V., O'Connor, F.M., Prather, M., Archibald, A., Brown, F., Deushi, M., Collins, W., Fiedler, S., Oshima, N., Murray, L.T., Samset, B.H., Smith, C., Turnock, S., Watson-Parris, D., Young, P.J., 2025. Opinion: The role of AerChemMIP in advancing climate and air quality research. Atmospheric Chemistry and Physics 25, 8289--8328. \url{https://doi.org/10.5194/acp-25-8289-2025}
\item Guenther, A.B., Jiang, X., Heald, C.L., Sakulyanontvittaya, T., Duhl, T., Emmons, L.K., Wang, X., 2012. The Model of Emissions of Gases and Aerosols from Nature version 2.1 (MEGAN2.1): an extended and updated framework for modeling biogenic emissions. Geoscientific Model Development 5, 1471--1492. \url{https://doi.org/10.5194/gmd-5-1471-2012}
\item Han, K., Wang, Y., Chen, H., Chen, X., Guo, J., Liu, Z., Tang, Y., Xiao, A., Xu, C., Xu, Y., Yang, Z., Zhang, Y., Tao, D., 2023. A Survey on Visual Transformer. IEEE Trans. Pattern Anal. Mach. Intell. 45, 87--110. \url{https://doi.org/10.1109/TPAMI.2022.3152247}
\item Henn, B., Bretherton, C.S., Koldunov, N., Lessig, C., Molina, M.J., Arcomano, T., Watt-Meyer, O., Couairon, G., Singh, R., Brunstein, R., Hasson, Y., Jost, A., Brenowitz, N., Manshausen, P., Cresswell-Clay, N., Durran, D., Hall, K.J.C., Yuval, J., Kochkov, D., Hoyer, S., Lopez-Gomez, I., 2026. AIMIP Phase 1: systematic evaluations of AI weather and climate models. \url{https://doi.org/10.48550/arXiv.2605.06944}
\item Hickman, S.H.M., Kelp, M.M., Griffiths, P.T., Doerksen, K., Miyazaki, K., Pennington, E.A., Koren, G., Iglesias-Suarez, F., Schultz, M.G., Chang, K.-L., Cooper, O.R., Archibald, A., Sommariva, R., Carlson, D., Wang, H., West, J.J., Liu, Z., 2025. Applications of Machine Learning and Artificial Intelligence in Tropospheric Ozone Research. Geoscientific Model Development 18, 8777--8800. \url{https://doi.org/10.5194/gmd-18-8777-2025}
\item Hoesly, R.M., Smith, S.J., Feng, L., Klimont, Z., Janssens-Maenhout, G., Pitkanen, T., Seibert, J.J., Vu, L., Andres, R.J., Bolt, R.M., Bond, T.C., Dawidowski, L., Kholod, N., Kurokawa, J., Li, M., Liu, L., Lu, Z., Moura, M.C.P., O'Rourke, P.R., Zhang, Q., 2018. Historical (1750--2014) anthropogenic emissions of reactive gases and aerosols from the Community Emissions Data System (CEDS). Geoscientific Model Development 11, 369--408. \url{https://doi.org/10.5194/gmd-11-369-2018}
\item Hourdin, F., Mauritsen, T., Gettelman, A., Golaz, J.-C., Balaji, V., Duan, Q., Folini, D., Ji, D., Klocke, D., Qian, Y., Rauser, F., Rio, C., Tomassini, L., Watanabe, M., Williamson, D., 2017. The Art and Science of Climate Model Tuning. Bulletin of the American Meteorological Society 98, 589--602. \url{https://doi.org/10.1175/BAMS-D-15-00135.1}
\item Hu, J.Y., Higuera-Mendieta, I., Sturm, P.O., Kelp, M.M., 2026. Toward Mechanistic Interpretability of an AI Foundation Model Fine-Tuned for Atmospheric Chemistry. \url{https://doi.org/10.48550/arXiv.2607.20778}
\item Inness, A., Ades, M., Agustí-Panareda, A., Barré, J., Benedictow, A., Blechschmidt, A.-M., Dominguez, J.J., Engelen, R., Eskes, H., Flemming, J., Huijnen, V., Jones, L., Kipling, Z., Massart, S., Parrington, M., Peuch, V.-H., Razinger, M., Remy, S., Schulz, M., Suttie, M., 2019. The CAMS reanalysis of atmospheric composition. Atmospheric Chemistry and Physics 19, 3515--3556. \url{https://doi.org/10.5194/acp-19-3515-2019}
\item Keller, C.A., Evans, M.J., 2019. Application of random forest regression to the calculation of gas-phase chemistry within the GEOS-Chem chemistry model v10. Geoscientific Model Development 12, 1209--1225. \url{https://doi.org/10.5194/gmd-12-1209-2019}
\item Kelp, M., Arzani, A., Castellanos, P., Griffiths, P., Higuera-Mendieta, I., Perez-Carrasco, M., Shah, V., Sturm, P.O., Weber, J., 2026. Strategic Governance of AI Models in Earth Science. \url{https://doi.org/10.48550/arXiv.2610.10560}
\item Kelp, M., Hickman, S., Miyazaki, K., Chang, K.-L., Griffiths, P., Zhu, Q., Koren, G., Iglesias-Suarez, F., Pennington, E., Schultz, M.G., 2025. AI4O3: A Foundational Data Collection for Artificial Intelligence in Tropospheric Ozone Research. Presented at the NeurIPS 2025 AI for Science Workshop.
\item Kelp, M.M., Jacob, D.J., Kutz, J.N., Marshall, J.D., Tessum, C.W., 2020. Toward Stable, General Machine-Learned Models of the Atmospheric Chemical System. Journal of Geophysical Research: Atmospheres 125, e2020JD032759. \url{https://doi.org/10.1029/2020JD032759}
\item Kelp, M.M., Jacob, D.J., Lin, H., Sulprizio, M.P., 2022. An Online-Learned Neural Network Chemical Solver for Stable Long-Term Global Simulations of Atmospheric Chemistry. Journal of Advances in Modeling Earth Systems 14, e2021MS002926. \url{https://doi.org/10.1029/2021MS002926}
\item Kelp, M.M., Tessum, C.W., Marshall, J.D., 2018. Orders-of-magnitude speedup in atmospheric chemistry modeling through neural network-based emulation. \url{https://doi.org/10.48550/arXiv.1808.03874}
\item Kirkpatrick, J., Pascanu, R., Rabinowitz, N., Veness, J., Desjardins, G., Rusu, A.A., Milan, K., Quan, J., Ramalho, T., Grabska-Barwinska, A., Hassabis, D., Clopath, C., Kumaran, D., Hadsell, R., 2017. Overcoming catastrophic forgetting in neural networks. Proceedings of the National Academy of Sciences 114, 3521--3526. \url{https://doi.org/10.1073/pnas.1611835114}
\item Kochkov, D., Yuval, J., Langmore, I., Norgaard, P., Smith, J., Mooers, G., Klöwer, M., Lottes, J., Rasp, S., Düben, P., Hatfield, S., Battaglia, P., Sanchez-Gonzalez, A., Willson, M., Brenner, M.P., Hoyer, S., 2024. Neural general circulation models for weather and climate. Nature 632, 1060--1066. \url{https://doi.org/10.1038/s41586-024-07744-y}
\item Lam, R., Sanchez-Gonzalez, A., Willson, M., Wirnsberger, P., Fortunato, M., Alet, F., Ravuri, S., Ewalds, T., Eaton-Rosen, Z., Hu, W., Merose, A., Hoyer, S., Holland, G., Vinyals, O., Stott, J., Pritzel, A., Mohamed, S., Battaglia, P., 2023. Learning skillful medium-range global weather forecasting. Science 382, 1416--1421. \url{https://doi.org/10.1126/science.adi2336}
\item Lehmann, F., Ozdemir, F., Cheng, Y., Hoefler, T., Schemm, S., Soja, B., Mishra, S., 2026. Can AI Weather Models Predict Beyond Two Weeks? A Quantitative Benchmark and Analysis of Long Rollouts. \url{https://doi.org/10.48550/arXiv.2605.30184}
\item Liu, Z., Li, K., Wild, O., Doherty, R.M., O'Connor, F.M., Turnock, S.T., 2025. Applying deep learning to a chemistry-climate model for improved ozone prediction. Atmospheric Chemistry and Physics 25, 16969--16981. \url{https://doi.org/10.5194/acp-25-16969-2025}
\item Long, M.S., Yantosca, R., Nielsen, J.E., Keller, C.A., da Silva, A., Sulprizio, M.P., Pawson, S., Jacob, D.J., 2015. Development of a grid-independent GEOS-Chem chemical transport model (v9-02) as an atmospheric chemistry module for Earth system models. Geoscientific Model Development 8, 595--602. \url{https://doi.org/10.5194/gmd-8-595-2015}
\item Miyazaki, K., Bowman, K.W., Yumimoto, K., Walker, T., Sudo, K., 2020. Evaluation of a multi-model, multi-constituent assimilation framework for tropospheric chemical reanalysis. Atmospheric Chemistry and Physics 20, 931--967. \url{https://doi.org/10.5194/acp-20-931-2020}
\item Miyazaki, K., Marchetti, Y., Montgomery, J., Lu, S., Bowman, K., 2025. Identifying drivers of surface ozone bias in global chemical reanalysis with explainable machine learning. Atmospheric Chemistry and Physics 25, 8507--8532. \url{https://doi.org/10.5194/acp-25-8507-2025}
\item Mulcahy, J.P., Jones, C.G., Rumbold, S.T., Kuhlbrodt, T., Dittus, A.J., Blockley, E.W., Yool, A., Walton, J., Hardacre, C., Andrews, T., Bodas-Salcedo, A., Stringer, M., de Mora, L., Harris, P., Hill, R., Kelley, D., Robertson, E., Tang, Y., 2023. UKESM1.1: development and evaluation of an updated configuration of the UK Earth System Model. Geoscientific Model Development 16, 1569--1600. \url{https://doi.org/10.5194/gmd-16-1569-2023}
\item Nguyen, T., Brandstetter, J., Kapoor, A., Gupta, J.K., Grover, A., 2023. ClimaX: A foundation model for weather and climate. \url{https://doi.org/10.48550/arXiv.2301.10343}
\item Pacifico, F., Harrison, S.P., Jones, C.D., Arneth, A., Sitch, S., Weedon, G.P., Barkley, M.P., Palmer, P.I., Serça, D., Potosnak, M., Fu, T.-M., Goldstein, A., Bai, J., Schurgers, G., 2011. Evaluation of a photosynthesis-based biogenic isoprene emission scheme in JULES and simulation of isoprene emissions under present-day climate conditions. Atmospheric Chemistry and Physics 11, 4371--4389. \url{https://doi.org/10.5194/acp-11-4371-2011}
\item Pfreundschuh, S., Kummerow, C.D., Schmude, J., Roy, S., Ramachandran, R., Lee, T., Anantharaj, V., Breen, K.H., 2026. Prithvi-Precip: Integrating Satellite Observations into an Atmospheric AI Foundation Model for Precipitation Forecasting. \url{https://doi.org/10.48550/arXiv.2608.03959}
\item Rasp, S., 2020. Coupled online learning as a way to tackle instabilities and biases in neural network parameterizations: general algorithms and Lorenz 96 case study (v1.0). Geoscientific Model Development 13, 2185--2196. \url{https://doi.org/10.5194/gmd-13-2185-2020}
\item Schmude, J., Roy, S., Trojak, W., Jakubik, J., Civitarese, D.S., Singh, S., Kuehnert, J., Ankur, K., Gupta, A., Phillips, C.E., Kienzler, R., Szwarcman, D., Gaur, V., Shinde, R., Lal, R., Silva, A.D., Diaz, J.L.G., Jones, A., Pfreundschuh, S., Lin, A., Sheshadri, A., Nair, U., Anantharaj, V., Hamann, H., Watson, C., Maskey, M., Lee, T.J., Moreno, J.B., Ramachandran, R., 2024. Prithvi WxC: Foundation Model for Weather and Climate. \url{https://doi.org/10.48550/arXiv.2409.13598}
\item Schneider, T., Lan, S., Stuart, A., Teixeira, J., 2017. Earth System Modeling 2.0: A Blueprint for Models That Learn From Observations and Targeted High-Resolution Simulations. Geophysical Research Letters 44, 12,396-12,417. \url{https://doi.org/10.1002/2017GL076101}
\item Sellar, A.A., Jones, C.G., Mulcahy, J.P., Tang, Y., Yool, A., Wiltshire, A., O'Connor, F.M., Stringer, M., Hill, R., Palmieri, J., Woodward, S., de Mora, L., Kuhlbrodt, T., Rumbold, S.T., Kelley, D.I., Ellis, R., Johnson, C.E., Walton, J., Abraham, N.L., Andrews, M.B., Andrews, T., Archibald, A.T., Berthou, S., Burke, E., Blockley, E., Carslaw, K., Dalvi, M., Edwards, J., Folberth, G.A., Gedney, N., Griffiths, P.T., Harper, A.B., Hendry, M.A., Hewitt, A.J., Johnson, B., Jones, A., Jones, C.D., Keeble, J., Liddicoat, S., Morgenstern, O., Parker, R.J., Predoi, V., Robertson, E., Siahaan, A., Smith, R.S., Swaminathan, R., Woodhouse, M.T., Zeng, G., Zerroukat, M., 2019. UKESM1: Description and Evaluation of the U.K. Earth System Model. Journal of Advances in Modeling Earth Systems 11, 4513--4558. \url{https://doi.org/10.1029/2019MS001739}
\item Sporre, M.K., Blichner, S.M., Schrödner, R., Karset, I.H.H., Berntsen, T.K., van Noije, T., Bergman, T., O'Donnell, D., Makkonen, R., 2020. Large difference in aerosol radiative effects from BVOC-SOA treatment in three Earth system models. Atmospheric Chemistry and Physics 20, 8953--8973. \url{https://doi.org/10.5194/acp-20-8953-2020}
\item Squire, O.J., Archibald, A.T., Griffiths, P.T., Jenkin, M.E., Smith, D., Pyle, J.A., 2015. Influence of isoprene chemical mechanism on modelled changes in tropospheric ozone due to climate and land use over the 21st century. Atmospheric Chemistry and Physics 15, 5123--5143. \url{https://doi.org/10.5194/acp-15-5123-2015}
\item Sturm, P.O., Wexler, A.S., 2022. Conservation laws in a neural network architecture: enforcing the atom balance of a Julia-based photochemical model (v0.2.0). Geoscientific Model Development 15, 3417--3431. \url{https://doi.org/10.5194/gmd-15-3417-2022}
\item Sturm, P.O., Wexler, A.S., 2020. A mass- and energy-conserving framework for using machine learning to speed computations: a photochemistry example. Geoscientific Model Development 13, 4435--4442. \url{https://doi.org/10.5194/gmd-13-4435-2020}
\item Telford, P.J., Abraham, N.L., Archibald, A.T., Braesicke, P., Dalvi, M., Morgenstern, O., O'Connor, F.M., Richards, N. a. D., Pyle, J.A., 2013. Implementation of the Fast-JX Photolysis scheme (v6.4) into the UKCA component of the MetUM chemistry-climate model (v7.3). Geoscientific Model Development 6, 161--177. \url{https://doi.org/10.5194/gmd-6-161-2013}
\item Thornhill, G., Collins, W., Olivié, D., Skeie, R.B., Archibald, A., Bauer, S., Checa-Garcia, R., Fiedler, S., Folberth, G., Gjermundsen, A., Horowitz, L., Lamarque, J.-F., Michou, M., Mulcahy, J., Nabat, P., Naik, V., O'Connor, F.M., Paulot, F., Schulz, M., Scott, C.E., Séférian, R., Smith, C., Takemura, T., Tilmes, S., Tsigaridis, K., Weber, J., 2021a. Climate-driven chemistry and aerosol feedbacks in CMIP6 Earth system models. Atmospheric Chemistry and Physics 21, 1105--1126. \url{https://doi.org/10.5194/acp-21-1105-2021}
\item Thornhill, G., Collins, W.J., Kramer, R.J., Olivié, D., Skeie, R.B., O'Connor, F.M., Abraham, N.L., Checa-Garcia, R., Bauer, S.E., Deushi, M., Emmons, L.K., Forster, P.M., Horowitz, L.W., Johnson, B., Keeble, J., Lamarque, J.-F., Michou, M., Mills, M.J., Mulcahy, J.P., Myhre, G., Nabat, P., Naik, V., Oshima, N., Schulz, M., Smith, C.J., Takemura, T., Tilmes, S., Wu, T., Zeng, G., Zhang, J., 2021b. Effective radiative forcing from emissions of reactive gases and aerosols -- a multi-model comparison. Atmospheric Chemistry and Physics 21, 853--874. \url{https://doi.org/10.5194/acp-21-853-2021}
\item Turnock, S.T., Reddington, C.L., West, J.J., O'Connor, F.M., 2023. The Air Pollution Human Health Burden in Different Future Scenarios That Involve the Mitigation of Near-Term Climate Forcers, Climate and Land-Use. GeoHealth 7, e2023GH000812. \url{https://doi.org/10.1029/2023GH000812}
\item Van Malderen, R., Thompson, A.M., Kollonige, D.E., Stauffer, R.M., Smit, H.G.J., Maillard Barras, E., Vigouroux, C., Petropavlovskikh, I., Leblanc, T., Thouret, V., Wolff, P., Effertz, P., Tarasick, D.W., Poyraz, D., Ancellet, G., De Backer, M.-R., Evan, S., Flood, V., Frey, M.M., Hannigan, J.W., Hernandez, J.L., Iarlori, M., Johnson, B.J., Jones, N., Kivi, R., Mahieu, E., McConville, G., Müller, K., Nagahama, T., Notholt, J., Piters, A., Prats, N., Querel, R., Smale, D., Steinbrecht, W., Strong, K., Sussmann, R., 2025. Global ground-based tropospheric ozone measurements: reference data and individual site trends (2000--2022) from the TOAR-II/HEGIFTOM project. Atmospheric Chemistry and Physics 25, 7187--7225. \url{https://doi.org/10.5194/acp-25-7187-2025}
\item Walters, D., Baran, A.J., Boutle, I., Brooks, M., Earnshaw, P., Edwards, J., Furtado, K., Hill, P., Lock, A., Manners, J., Morcrette, C., Mulcahy, J., Sanchez, C., Smith, C., Stratton, R., Tennant, W., Tomassini, L., Van Weverberg, K., Vosper, S., Willett, M., Browse, J., Bushell, A., Carslaw, K., Dalvi, M., Essery, R., Gedney, N., Hardiman, S., Johnson, B., Johnson, C., Jones, A., Jones, C., Mann, G., Milton, S., Rumbold, H., Sellar, A., Ujiie, M., Whitall, M., Williams, K., Zerroukat, M., 2019. The Met Office Unified Model Global Atmosphere 7.0/7.1 and JULES Global Land 7.0 configurations. Geoscientific Model Development 12, 1909--1963. \url{https://doi.org/10.5194/gmd-12-1909-2019}
\item Watt-Meyer, O., Dresdner, G., McGibbon, J., Clark, S.K., Henn, B., Duncan, J., Brenowitz, N.D., Kashinath, K., Pritchard, M.S., Bonev, B., Peters, M.E., Bretherton, C.S., 2023. ACE: A fast, skillful learned global atmospheric model for climate prediction. \url{https://doi.org/10.48550/arXiv.2310.02074}
\item Watt-Meyer, O., Henn, B., McGibbon, J., Clark, S.K., Kwa, A., Perkins, W.A., Wu, E., Harris, L., Bretherton, C.S., 2025. ACE2: accurately learning subseasonal to decadal atmospheric variability and forced responses. npj Clim Atmos Sci 8, 205. \url{https://doi.org/10.1038/s41612-025-01090-0}
\item Weber, J., Archer-Nicholls, S., Abraham, N.L., Shin, Y.M., Bannan, T.J., Percival, C.J., Bacak, A., Artaxo, P., Jenkin, M., Khan, M.A.H., Shallcross, D.E., Schwantes, R.H., Williams, J., Archibald, A.T., 2021. Improvements to the representation of BVOC chemistry--climate interactions in UKCA (v11.5) with the CRI-Strat 2 mechanism: incorporation and evaluation. Geoscientific Model Development 14, 5239--5268. \url{https://doi.org/10.5194/gmd-14-5239-2021}
\item Weber, J., King, J.A., Abraham, N.L., Grosvenor, D.P., Smith, C.J., Shin, Y.M., Lawrence, P., Roe, S., Beerling, D.J., Martin, M.V., 2024. Chemistry-albedo feedbacks offset up to a third of forestation's CO2 removal benefits. Science 383, 860--864. \url{https://doi.org/10.1126/science.adg6196}
\item Weber, J., Wright, M., Collins, W.J., Shine, K.P., O'Connor, F.M., Folberth, G.A., Griffiths, P., Abernethy, S., 2026. Methane Emission Reductions Slow Stratospheric Ozone Recovery by Amplifying the Potency of Ozone Depleting Substances. Geophysical Research Letters 53, e2025GL119900. \url{https://doi.org/10.1029/2025GL119900}
\item Wilcox, L.J., Allen, R.J., Samset, B.H., Bollasina, M.A., Griffiths, P.T., Keeble, J., Lund, M.T., Makkonen, R., Merikanto, J., O'Donnell, D., Paynter, D.J., Persad, G.G., Rumbold, S.T., Takemura, T., Tsigaridis, K., Undorf, S., Westervelt, D.M., 2023. The Regional Aerosol Model Intercomparison Project (RAMIP). Geoscientific Model Development 16, 4451--4479. \url{https://doi.org/10.5194/gmd-16-4451-2023}
\item Wild, O., Ryan, E.M., 2026. Quantifying and addressing the uncertainties in tropospheric ozone and OH in a global chemistry transport model. Atmospheric Chemistry and Physics 26, 2255--2273. \url{https://doi.org/10.5194/acp-26-2255-2026}
\item Wild, O., Voulgarakis, A., O'Connor, F., Lamarque, J.-F., Ryan, E.M., Lee, L., 2020. Global sensitivity analysis of chemistry--climate model budgets of tropospheric ozone and OH: exploring model diversity. Atmospheric Chemistry and Physics 20, 4047--4058. \url{https://doi.org/10.5194/acp-20-4047-2020}
\item Xia, Z., Zhao, C., Du, Q., Yang, Z., Zhang, M., Qiao, L., 2025. Advancing Sophisticated Photochemistry Simulation in Atmospheric Numerical Models With Artificial Intelligence PhotoChemistry (AIPC) Scheme Using the Feature-Mapping Subspace Self-Attention Algorithm. Journal of Advances in Modeling Earth Systems 17, e2024MS004646. \url{https://doi.org/10.1029/2024MS004646}
\end{list}

\clearpage
\setcounter{figure}{0}\setcounter{table}{0}\setcounter{equation}{0}
\renewcommand{\thefigure}{S\arabic{figure}}
\renewcommand{\thetable}{S\arabic{table}}
\renewcommand{\theequation}{S\arabic{equation}}
\subsection*{Text S1. Per-species temporal weighting}
We apply a second, complementary per-species weighting derived from the temporal autocorrelation of each species' global mean concentration in the training record. This loss is meant to give more weight to species whose single-step errors are propagated furthest through an autoregressive rollout as an error in a slowly evolving field can accumulate into drift. For each species and pressure level, we compute the global mean concentration on every training day, correlate that series with itself at lags of 1 to 365 days, and fit an exponential to the result. Writing τ(v) for the median of the fitted timescale across pressure levels and $\tilde{\tau}$ for its median across species, each species is weighted by:

\begin{equation}
w_{\tau}(v) = \min\left[\max\left(\frac{\tau(v)}{\tilde{\tau}},\,0.5\right),\,5.0\right]
\end{equation}

The resulting weights are then divided by their mean across species, so that this term redistributes weight between species without changing the overall magnitude of the chemistry loss. The final weights span 0.38 to 3.04, and only the lower bound of Eq. S1 is active. We find that these global mean series are dominated by the seasonal cycle, such that the fitted timescales order species by the regularity of their cycle rather than by their chemical lifetime. Therefore, this term is best understood as a moderate empirical reweighting rather than a chemical lifetime-based prior.

\subsection*{Text S2. Optimization settings}
We use standard optimization settings. Training uses a learning rate of 1e-4 with an Adam optimizer and a plateau learning rate schedule with patience of 3 epochs, a decay factor of 0.5, and a learning rate minimum of 1e-6. We apply gradient clipping at 1.0 and early stopping after 8 epochs without validation improvement for up to 200 epochs total.

\subsection*{Text S3. Evaluation metrics}
We report several standard error metrics for both single-step and rollout evaluation. Root mean square error (RMSE) summarizes the magnitude of prediction error and normalized mean bias (NMB) summarizes its systematic sign and size relative to the target:

\begin{equation}
\mathrm{RMSE}(v,l) = \sqrt{\frac{1}{N}\sum_{i=1}^{N}\left(\mathrm{pred}_{i}(v,l) - \mathrm{target}_{i}(v,l)\right)^{2}}
\end{equation}

\begin{equation}
\mathrm{NMB}(v,l) = \frac{\sum_{i=1}^{N}\left(\mathrm{pred}_{i}(v,l) - \mathrm{target}_{i}(v,l)\right)}{\sum_{i=1}^{N}\mathrm{target}_{i}(v,l)} \times 100\%
\end{equation}

The coefficient of determination summarizes how much of the target's variance the prediction explains:

\begin{equation}
R^{2}(v,l) = 1 - \frac{\sum_{i=1}^{N}\left(\mathrm{pred}_{i}(v,l) - \mathrm{target}_{i}(v,l)\right)^{2}}{\sum_{i=1}^{N}\left(\mathrm{target}_{i}(v,l) - \underline{\mathrm{target}}(v,l)\right)^{2}}, \qquad \underline{\mathrm{target}}(v,l) = \frac{1}{N}\sum_{i=1}^{N}\mathrm{target}_{i}(v,l)
\end{equation}

\emph{N} is the total number of evaluation points for variable \emph{v} at level \emph{l}, pooled across grid cells for single-step metrics and additionally across every rollout time step for rollout metrics. All pooled metrics are weighted by the cosine of the latitude for area and by layer pressure thickness for mass, so that each species is summarized by one globally representative value rather than one dominated by thin upper levels. Levels above each species' prescribed ceiling (Section 2.3) are excluded.

\clearpage
\begin{table}[htbp]
\centering
\caption{Single step tropospheric (1000-200 hPa) prediction skill of Aurora\textsubscript{UKCA} against UKCA.\textsuperscript{a,b}}
\label{tab:S1}
\small
\begin{tabular}{@{}llrrr@{}}
\toprule
Field & Units & R\textsuperscript{2} & NMB (\%) & RMSE \\
\midrule
\multicolumn{5}{@{}l}{\textit{Gas phase chemistry}} \\
O₃ & ppb & 0.98 & +0.75 & 5.77 \\
OH & ppt & 0.95 & +1.95 & 0.018 \\
HO₂ & ppt & 0.97 & +1.54 & 0.61 \\
HNO₃ & ppb & 0.90 & +5.46 & 0.16 \\
CO & ppb & 0.88 & −0.66 & 10.7 \\
CH₄ & ppb & 0.96 & −0.51 & 32.1 \\
HCHO & ppb & 0.97 & +3.35 & 0.064 \\
PAN & ppb & 0.91 & +6.05 & 0.053 \\
Isoprene & ppb & 0.96 & −2.70 & 0.085 \\
NOₓ & ppb & 0.76 & −4.60 & 0.21 \\
SO₂ & ppb & 0.68 & −1.78 & 0.28 \\
ClOₓ & ppt & 0.82 & +1.47 & 1.16 \\
ClONO₂ & ppb & 0.95 & +1.05 & 1.51×10\textsuperscript{−3} \\
HCl & ppb & 0.96 & +0.69 & 5.99×10\textsuperscript{−3} \\
MeBr & ppt & 0.89 & +2.43 & 0.57 \\
Cl\_precursor & ppb & 0.96 & +0.35 & 0.021 \\
\midrule
\multicolumn{5}{@{}l}{\textit{Aerosol}} \\
BC aerosol & kg kg\textsuperscript{−1} & 0.90 & −6.75 & 3.57×10\textsuperscript{−11} \\
SO₄ aerosol & kg kg\textsuperscript{−1} & 0.84 & −6.64 & 2.90×10\textsuperscript{−10} \\
\midrule
\multicolumn{5}{@{}l}{\textit{Radiation}} \\
SW downwelling & W m\textsuperscript{−2} & 0.98 & +1.63 & 16.8 \\
SW upwelling & W m\textsuperscript{−2} & 0.92 & +1.03 & 14.9 \\
LW downwelling & W m\textsuperscript{−2} & 0.99 & −0.39 & 8.58 \\
LW upwelling & W m\textsuperscript{−2} & 0.99 & −0.33 & 5.79 \\
\midrule
\multicolumn{5}{@{}l}{\textit{Meteorology}} \\
Temperature & K & 0.97 & −0.16 & 4.90 \\
Zonal wind & m s\textsuperscript{−1} & 0.98 & +1.98 & 1.77 \\
Meridional wind & m s\textsuperscript{−1} & 0.96 & ---\textsuperscript{c} & 1.65 \\
Specific humidity & g kg\textsuperscript{−1} & 0.99 & −1.74 & 0.44 \\
Geopotential height & m & 1.000 & −0.14 & 30.1 \\
\bottomrule
\end{tabular}

\vspace{4pt}
\begin{minipage}{\textwidth}\footnotesize
\textsuperscript{a} Metrics are pooled over the 3D grid in the troposphere weighted by the cosine of latitude to account for grid cell area and by layer pressure thickness to account for atmospheric mass.\\
\textsuperscript{b} Evaluated over the 358 daily steps of the held out 2014 test year. Each prediction is a single forward pass initialised from UKCA with no autoregressive chaining.\\
\textsuperscript{c} NMB is not reported because the regional mean is close to zero, which makes the normalised ratio unstable. R\textsuperscript{2} and RMSE remain valid.
\end{minipage}
\end{table}
\begin{table}[htbp]
\centering
\caption{Single step stratospheric (150-1 hPa) prediction skill of Aurora\textsubscript{UKCA} against UKCA.\textsuperscript{a,b,c}}
\label{tab:S2}
\small
\begin{tabular}{@{}llrrr@{}}
\toprule
Field & Units & R\textsuperscript{2} & NMB (\%) & RMSE \\
\midrule
\multicolumn{5}{@{}l}{\textit{Gas phase chemistry}} \\
O₃ & ppb & 0.99 & +0.47 & 129 \\
OH & ppt & 1.000 & +0.80 & 0.46 \\
HO₂ & ppt & 0.99 & +2.30 & 0.73 \\
HNO₃ & ppb & 0.98 & −5.16 & 0.30 \\
CO & ppb & 0.88 & −3.25 & 8.92 \\
CH₄ & ppb & 0.99 & −0.61 & 35.6 \\
HCHO & ppb & 0.89 & −2.16 & 0.019 \\
NOₓ & ppb & 0.99 & −4.61 & 0.21 \\
SO₂ & ppb & 0.70 & −10.66 & 0.014 \\
ClOₓ & ppt & 0.96 & +4.24 & 17.5 \\
ClONO₂ & ppb & 0.98 & +1.53 & 0.051 \\
HCl & ppb & 0.99 & +0.56 & 0.062 \\
MeBr\textsuperscript{d} & ppt & 0.99 & +1.97 & 0.56 \\
Cl\_precursor & ppb & 0.99 & +0.44 & 0.027 \\
\midrule
\multicolumn{5}{@{}l}{\textit{Radiation}} \\
SW downwelling & W m\textsuperscript{−2} & 0.99 & +1.39 & 6.52 \\
SW upwelling & W m\textsuperscript{−2} & 0.91 & +1.52 & 18.1 \\
LW downwelling & W m\textsuperscript{−2} & 0.97 & −0.90 & 0.95 \\
LW upwelling & W m\textsuperscript{−2} & 0.95 & −0.52 & 10.6 \\
\midrule
\multicolumn{5}{@{}l}{\textit{Meteorology}} \\
Temperature & K & 0.99 & −0.25 & 1.26 \\
Zonal wind & m s\textsuperscript{−1} & 0.98 & −0.25 & 2.95 \\
Meridional wind & m s\textsuperscript{−1} & 0.93 & ---\textsuperscript{e} & 2.05 \\
Specific humidity & g kg\textsuperscript{−1} & 0.92 & −2.74 & 9.13×10\textsuperscript{−4} \\
Geopotential height & m & 1.000 & −0.19 & 53.5 \\
\bottomrule
\end{tabular}

\vspace{4pt}
\begin{minipage}{\textwidth}\footnotesize
\textsuperscript{a} Metrics are pooled over the 3D grid in the stratosphere and weighted by the cosine of latitude to account for grid cell area and by layer pressure thickness to account for atmospheric mass.\\
\textsuperscript{b} Evaluated over the 358 daily steps of the held out 2014 test year. Each prediction is a single forward pass initialised from UKCA with no autoregressive chaining.\\
\textsuperscript{c} PAN, Isoprene, BC aerosol, and sulfate aerosol are set to zero above 150 hPa during training. These species are excluded from the stratospheric results.\\
\textsuperscript{d} Set to zero above 10 hPa during training configuration. These levels are excluded.\\
\textsuperscript{e} NMB is not reported because the regional mean is close to zero, which makes the normalised ratio unstable. R\textsuperscript{2} and RMSE remain valid.
\end{minipage}
\end{table}
\clearpage

\begin{figure}[htbp]
\centering
\includegraphics[width=\textwidth,height=0.82\textheight,keepaspectratio]{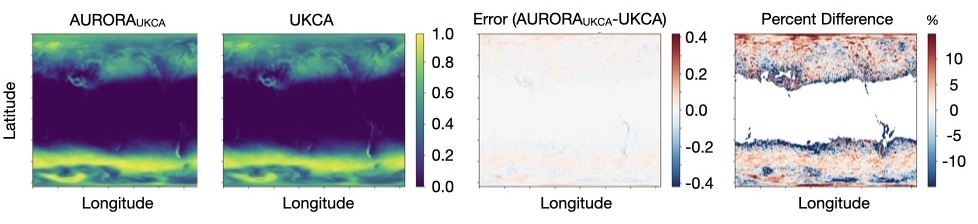}
\caption{\textbf{Annual mean low-cloud fraction from 1000 to 700 hPa.}}
\label{fig:S1}
\end{figure}

\begin{figure}[htbp]
\centering
\includegraphics[width=\textwidth,height=0.82\textheight,keepaspectratio]{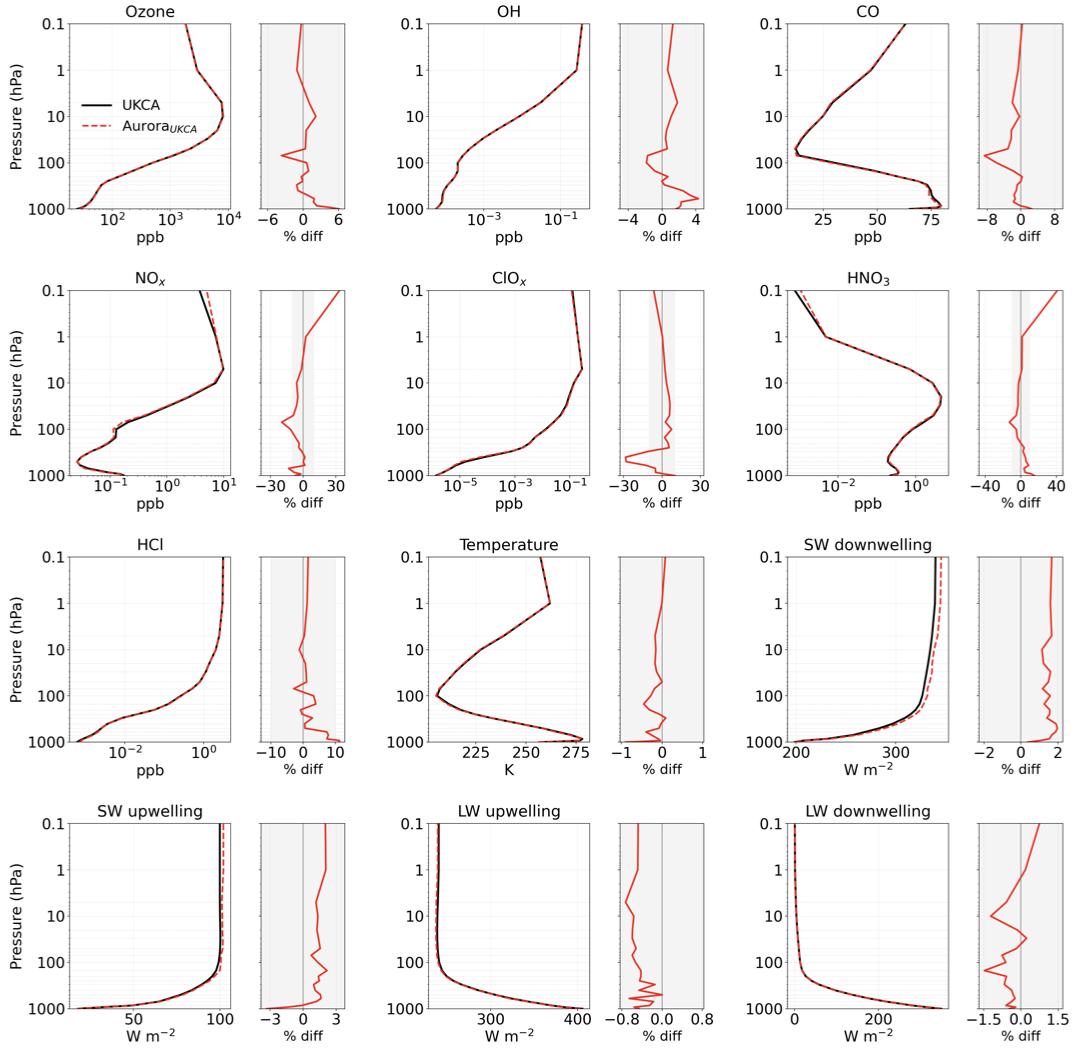}
\caption{\textbf{Annual global mean vertical profiles}. Each pair of panels shows the profile on the left and the percent difference on the right. Black lines are UKCA and red dashed lines are Aurora\textsubscript{UKCA}. The grey band marks ±10\%. Profiles are area-weighted global means accumulated over 358 single step predictions spanning the 2014 test year. There is no autoregressive rollout, so every prediction is initialized from UKCA. Pressure is logarithmic from the surface to 0.1 hPa and species spanning orders of magnitude are plotted on a logarithmic horizontal axis.}
\label{fig:S2}
\end{figure}

\begin{figure}[htbp]
\centering
\includegraphics[width=\textwidth,height=0.82\textheight,keepaspectratio]{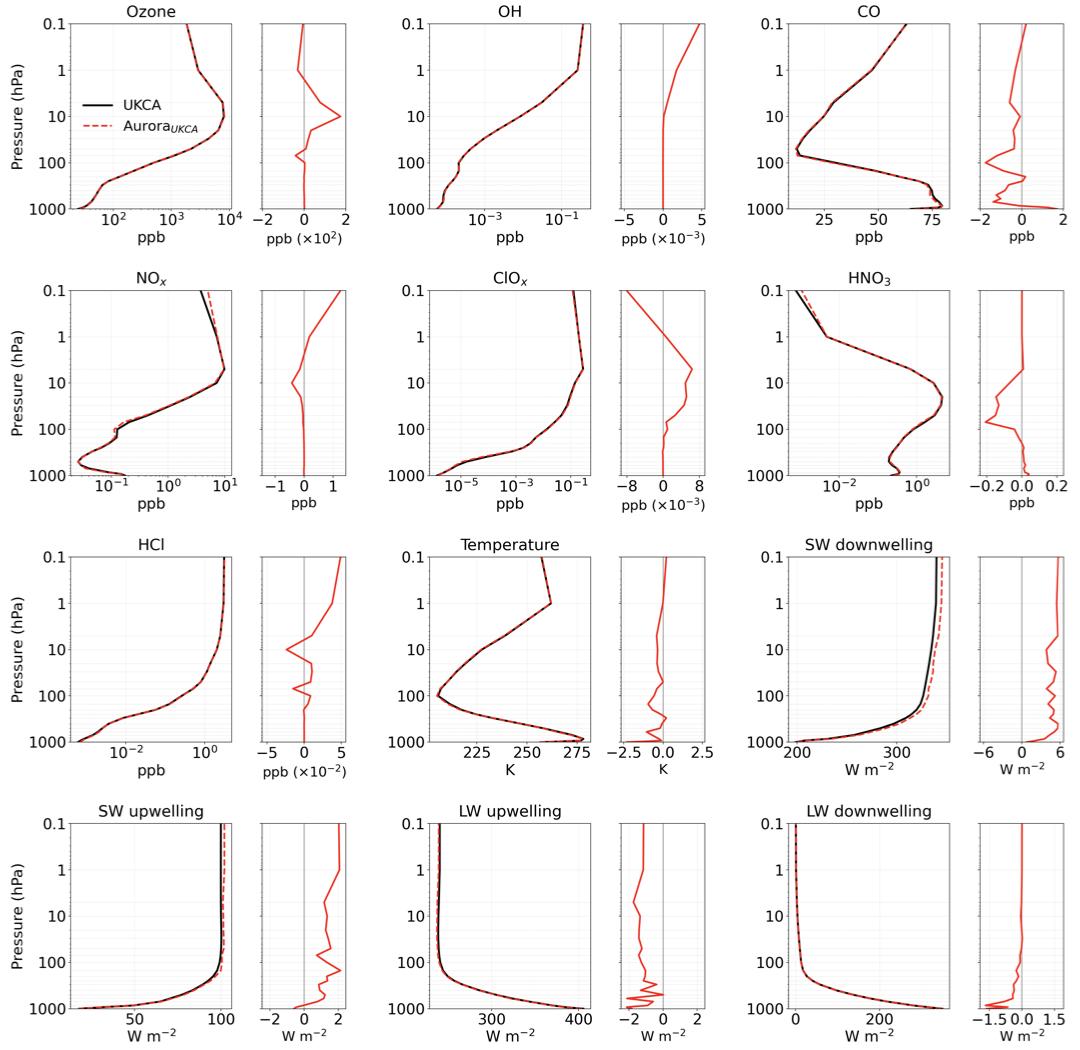}
\caption{\textbf{Annual global mean vertical profiles as in Figure S2, but with absolute differences.}}
\label{fig:S3}
\end{figure}

\begin{figure}[htbp]
\centering
\includegraphics[width=\textwidth,height=0.82\textheight,keepaspectratio]{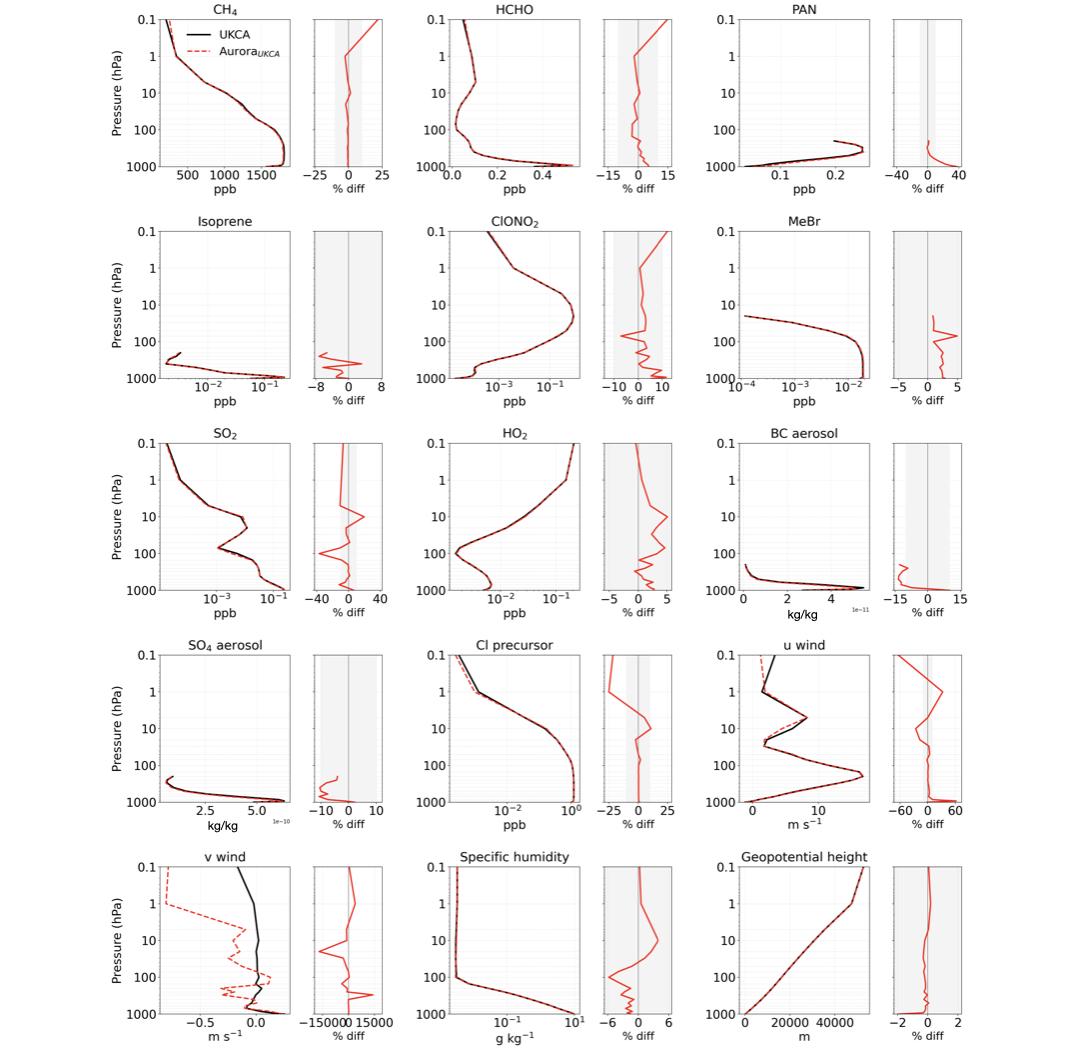}
\caption{\textbf{Annual global mean vertical profiles as in Figure S2, but with remaining variables.}}
\label{fig:S4}
\end{figure}

\begin{figure}[htbp]
\centering
\includegraphics[width=\textwidth,height=0.82\textheight,keepaspectratio]{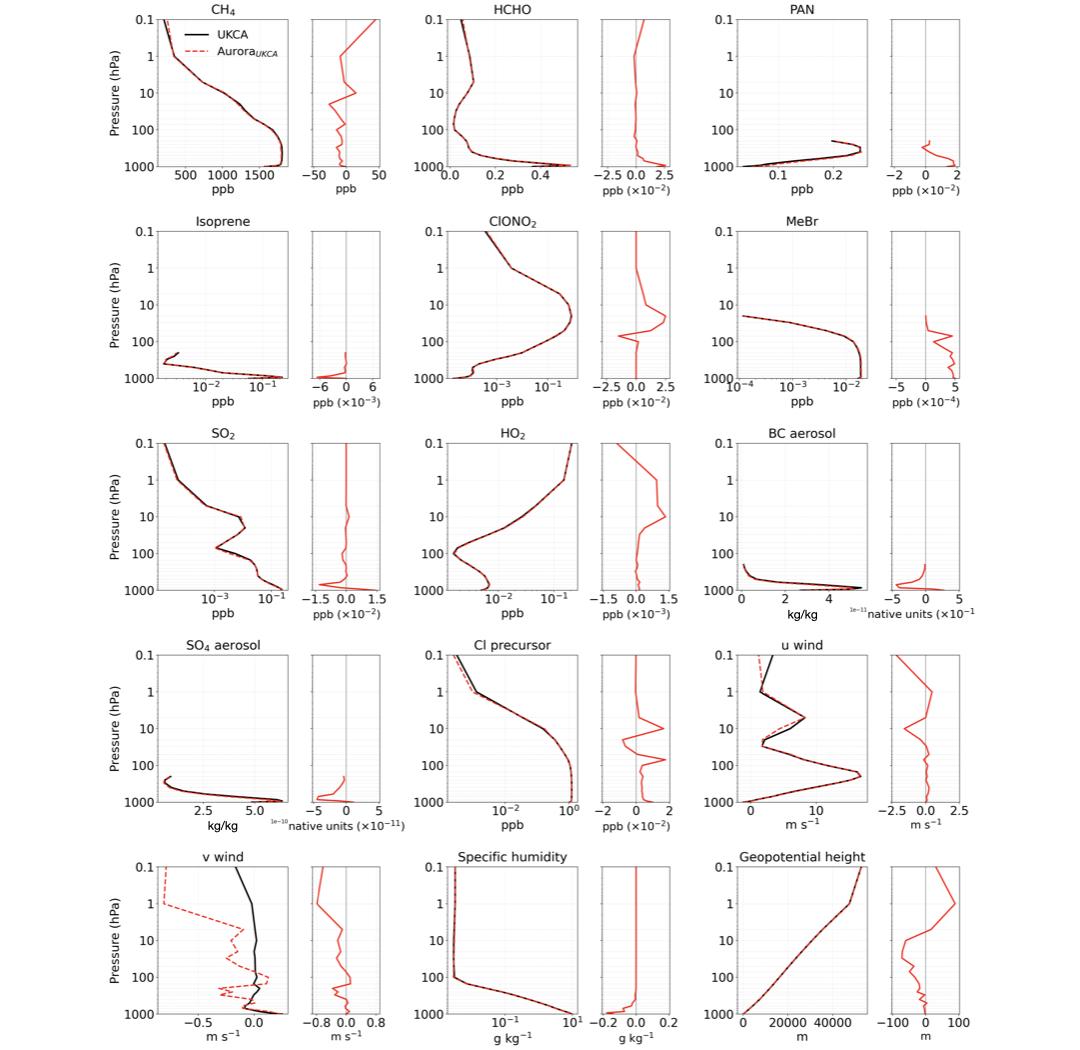}
\caption{\textbf{Annual global mean vertical profiles as in Figure S4, but with absolute differences.}}
\label{fig:S5}
\end{figure}

\begin{figure}[htbp]
\centering
\includegraphics[width=\textwidth,height=0.82\textheight,keepaspectratio]{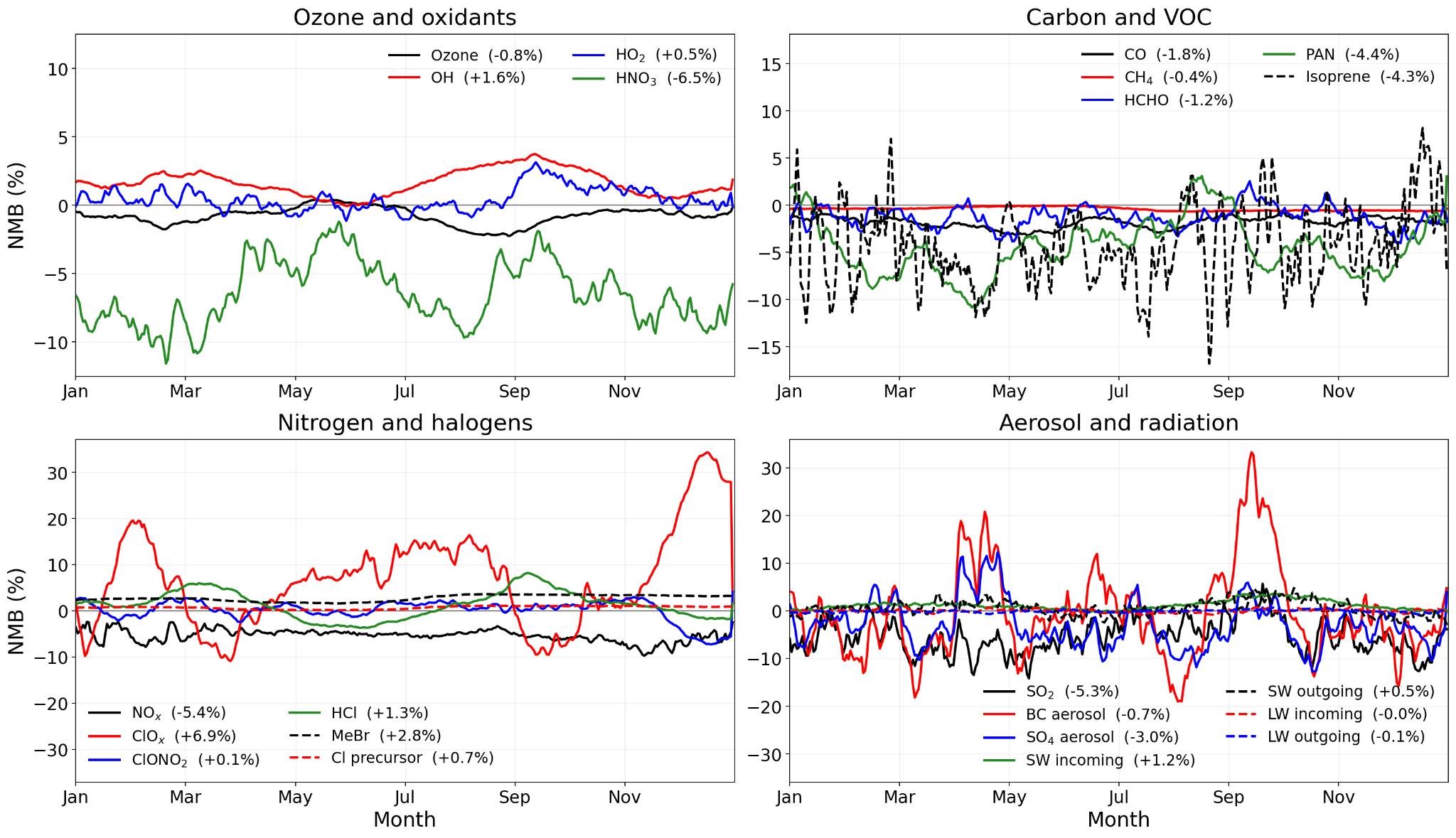}
\caption{\textbf{Year-long atmospheric column evolution of the global normalized mean bias (NMB) over a Aurora\textsubscript{UKCA} simulation, averaged over five annual cycles of the free-running rollout.} Each field's annual mean is given in the legend.}
\label{fig:S6}
\end{figure}

\begin{figure}[htbp]
\centering
\includegraphics[width=\textwidth,height=0.82\textheight,keepaspectratio]{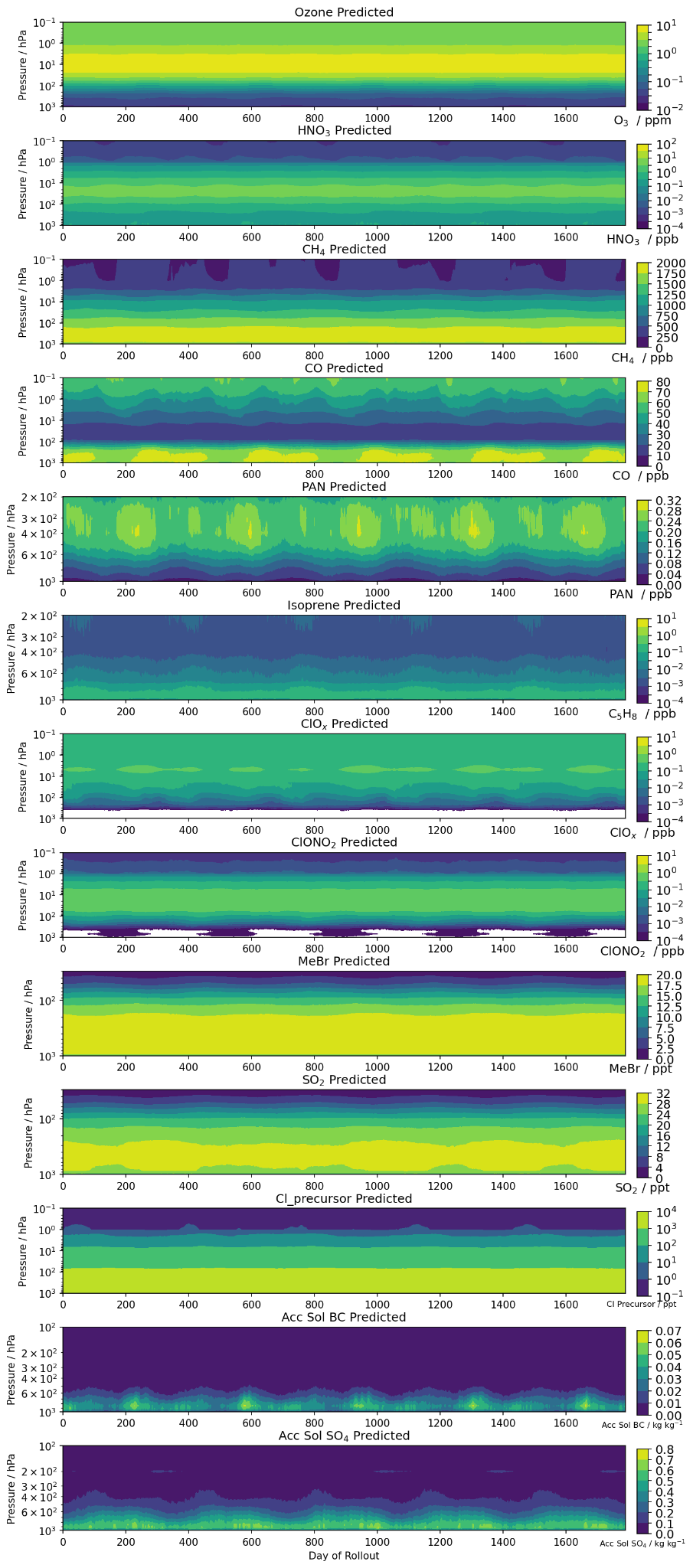}
\caption{\textbf{Global mean vertical structure of predicted composition over the five year Aurora\textsubscript{UKCA} rollout.}}
\label{fig:S7}
\end{figure}

\begin{figure}[htbp]
\centering
\includegraphics[width=\textwidth,height=0.82\textheight,keepaspectratio]{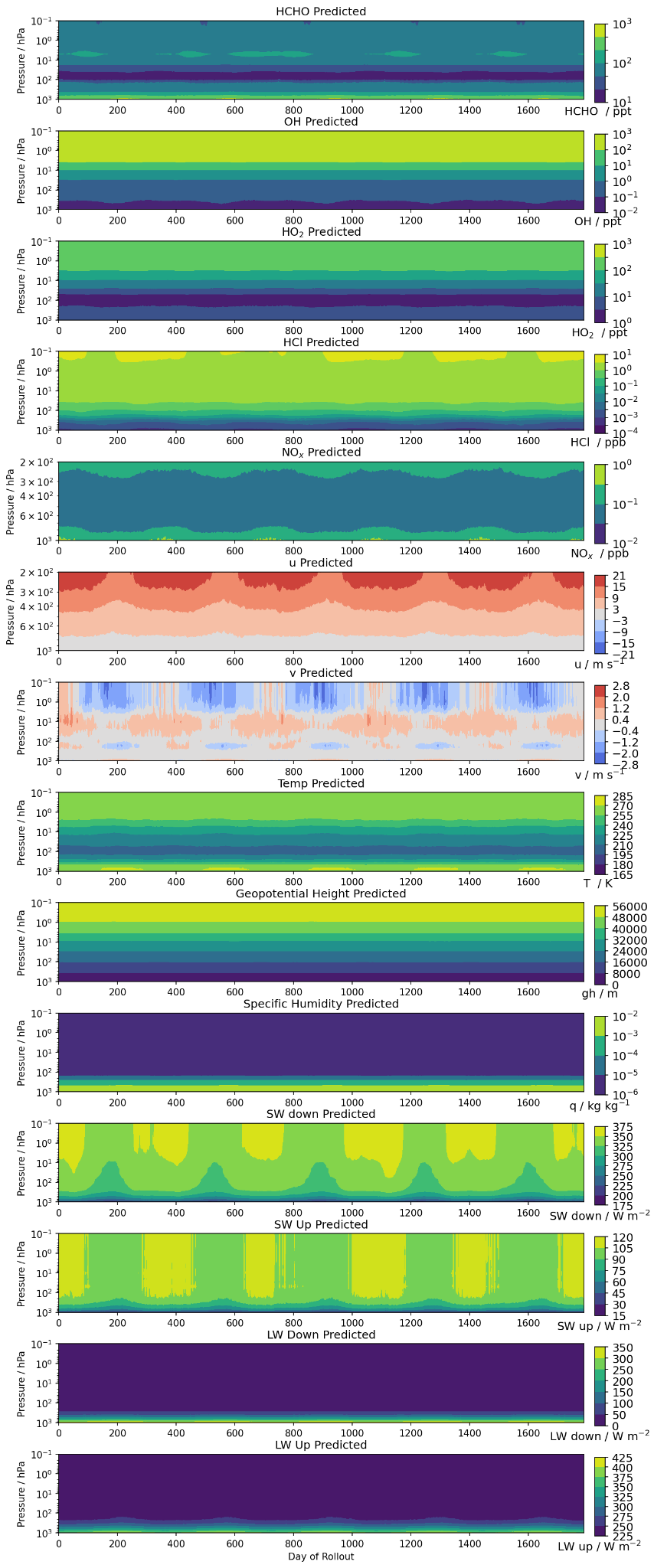}
\caption{\textbf{As in Figure S7, but for the remaining variables.}}
\label{fig:S8}
\end{figure}

\begin{figure}[htbp]
\centering
\includegraphics[width=\textwidth,height=0.82\textheight,keepaspectratio]{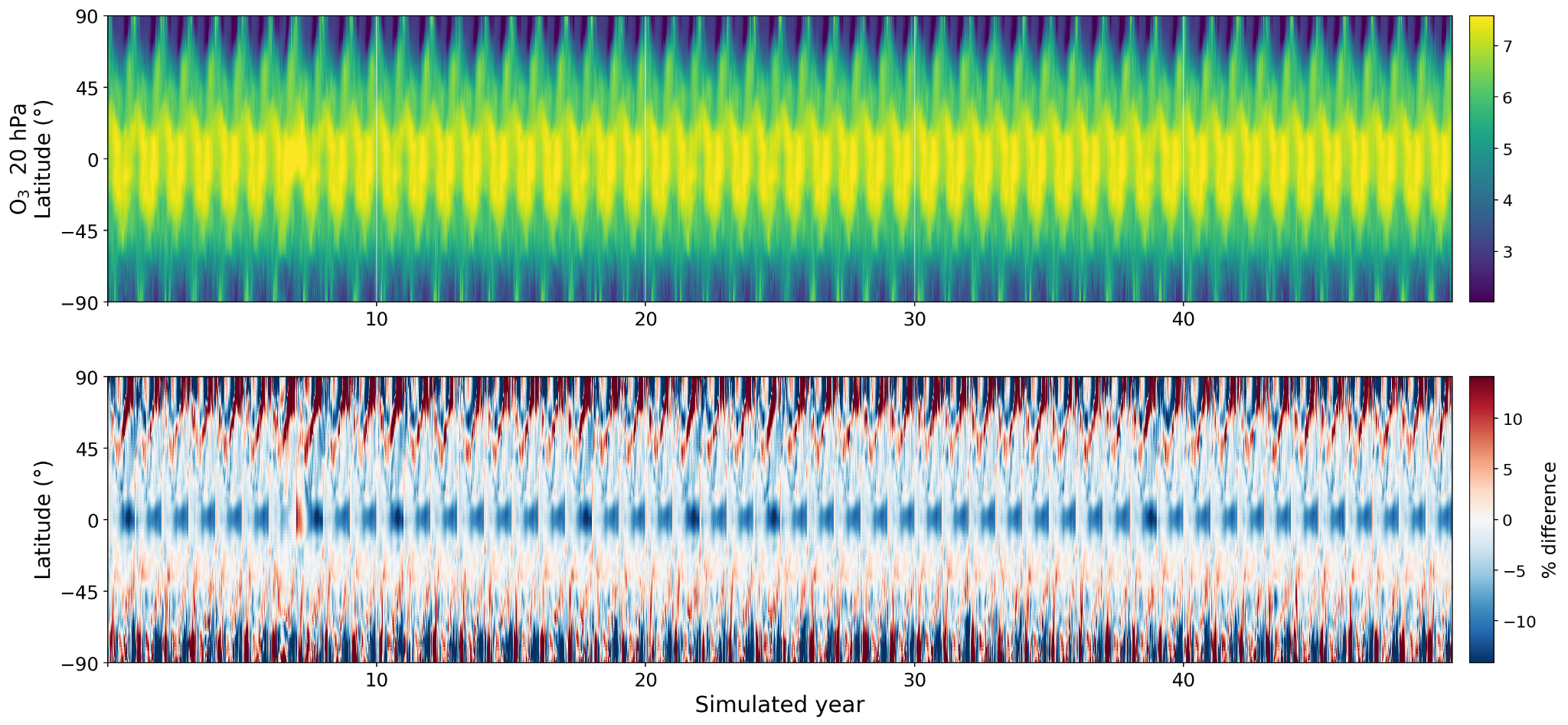}
\caption{\textbf{50-year rollout for ozone at 20 hPa.}}
\label{fig:S9}
\end{figure}

\begin{figure}[htbp]
\centering
\includegraphics[width=\textwidth,height=0.82\textheight,keepaspectratio]{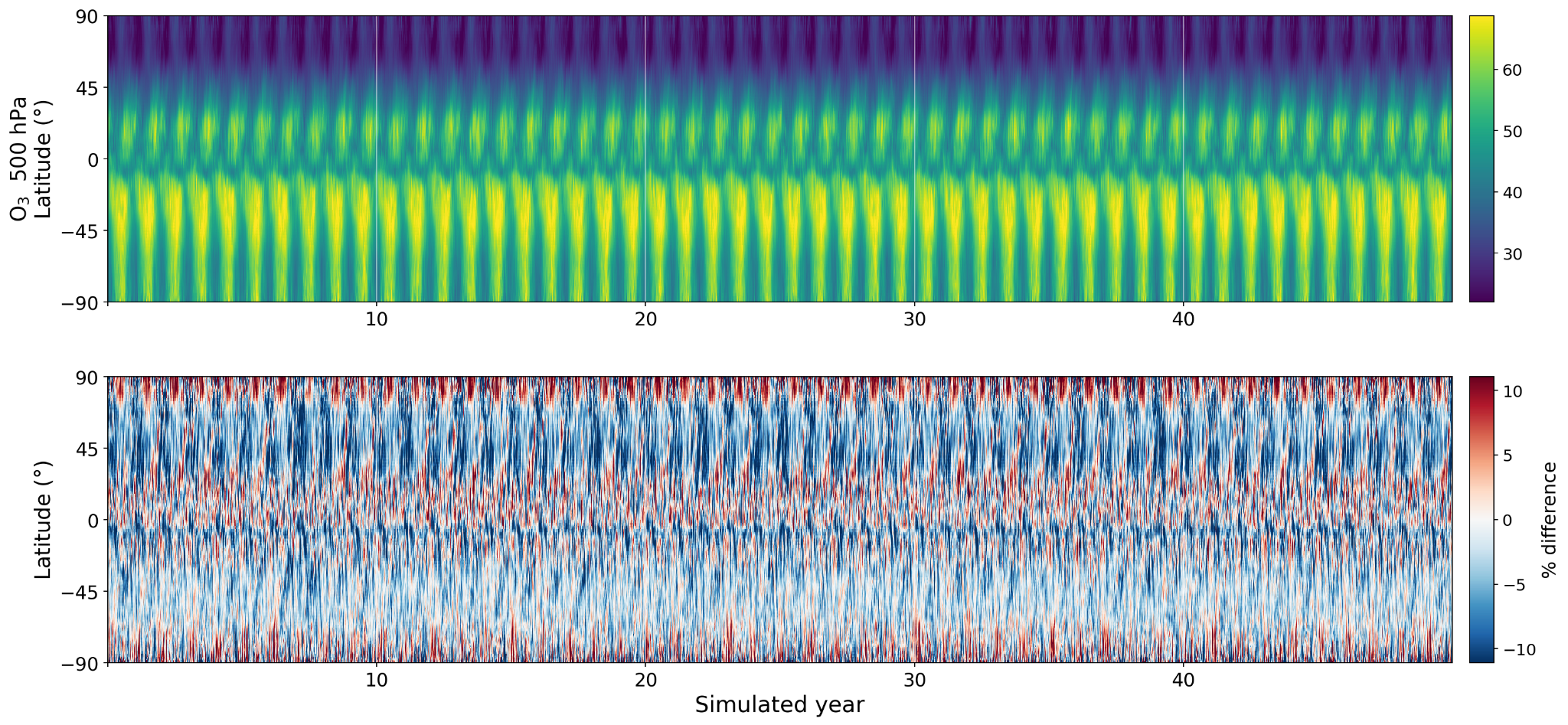}
\caption{\textbf{50-year rollout for ozone at 500 hPa.}}
\label{fig:S10}
\end{figure}

\begin{figure}[htbp]
\centering
\includegraphics[width=\textwidth,height=0.82\textheight,keepaspectratio]{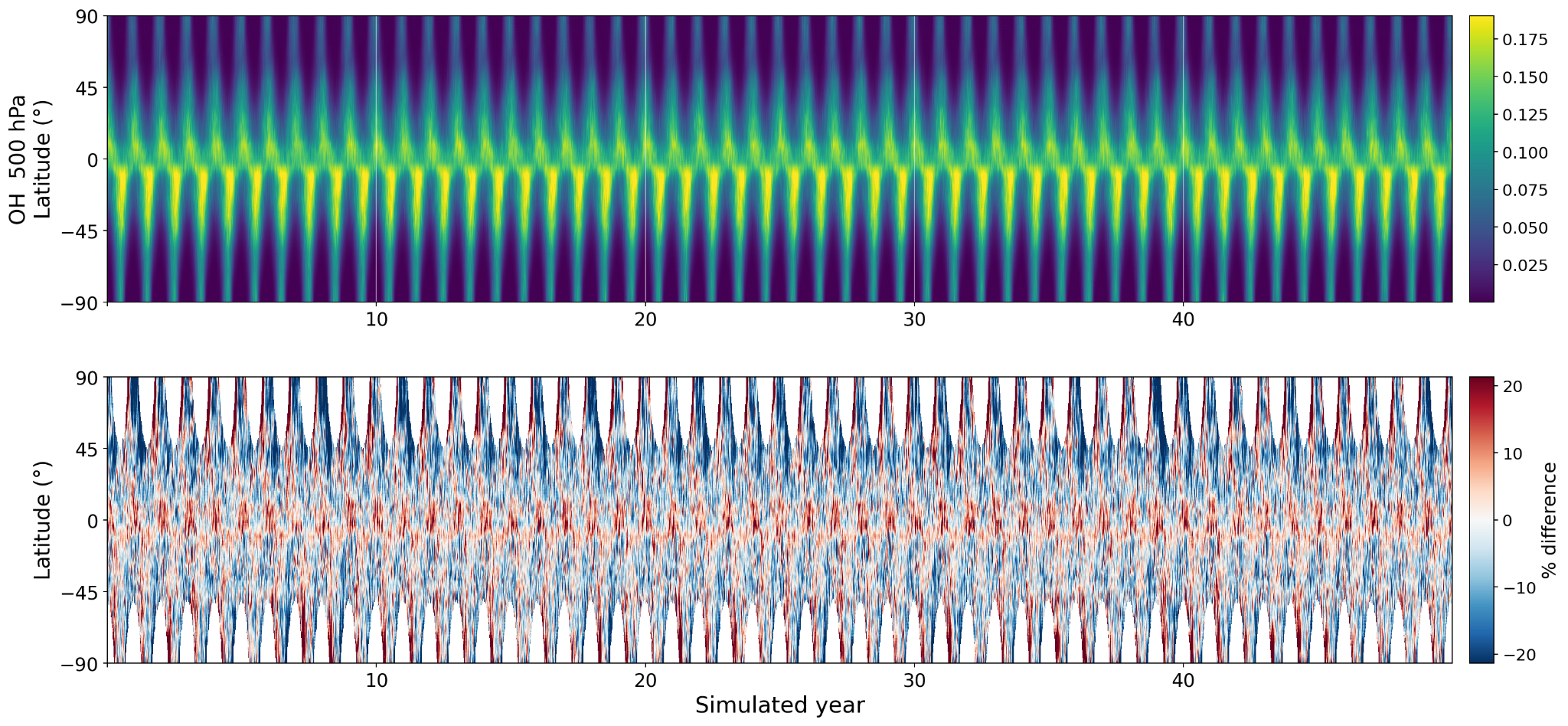}
\caption{\textbf{50-year rollout for OH at 500 hPa.}}
\label{fig:S11}
\end{figure}

\begin{figure}[htbp]
\centering
\includegraphics[width=\textwidth,height=0.82\textheight,keepaspectratio]{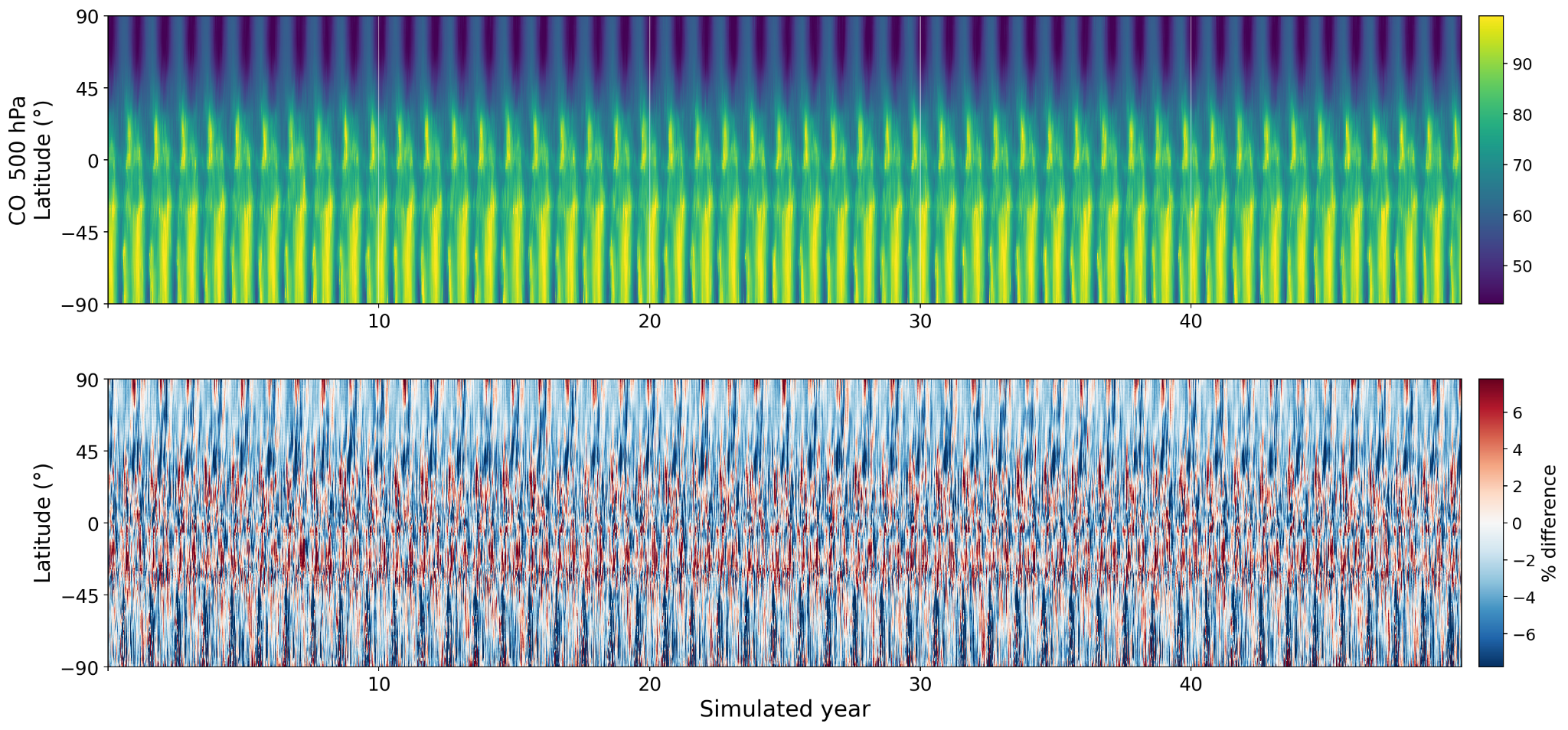}
\caption{\textbf{50-year rollout for CO at 500 hPa.}}
\label{fig:S12}
\end{figure}

\begin{figure}[htbp]
\centering
\includegraphics[width=\textwidth,height=0.82\textheight,keepaspectratio]{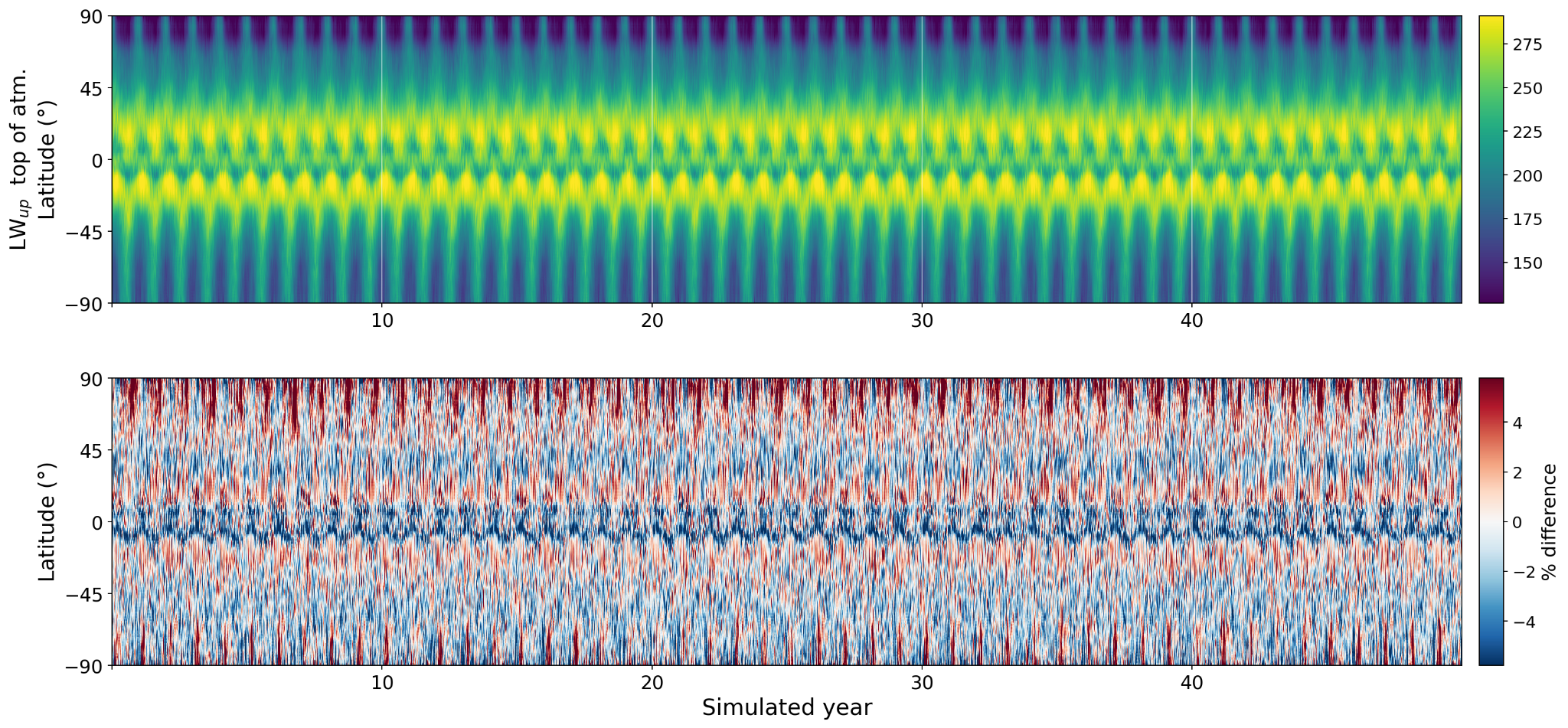}
\caption{\textbf{50-year rollout for upwelling LW radiation at top of atmosphere.}}
\label{fig:S13}
\end{figure}

\begin{figure}[htbp]
\centering
\includegraphics[width=\textwidth,height=0.82\textheight,keepaspectratio]{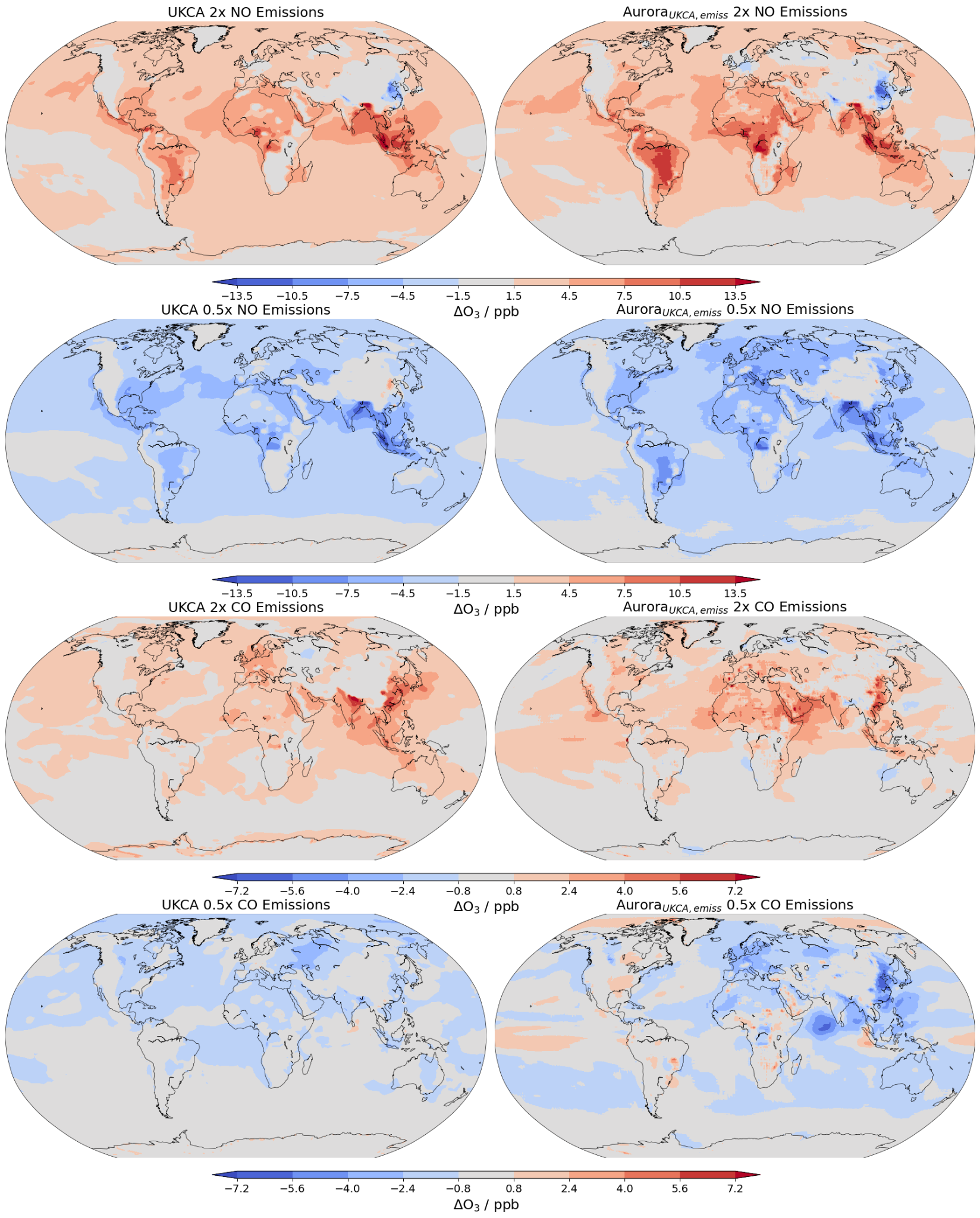}
\caption{\textbf{Ozone responses at 925 hPa to emission scalings used in training.} Change in annual mean ozone mole fraction at 925 hPa relative to the control following scalings of 2 to NO emissions (first row), 0.5 to NO emissions (second row), 2 to both CO emissions (third row), and 0.5 to CO emissions (fourth row).}
\label{fig:S14}
\end{figure}

\begin{figure}[htbp]
\centering
\includegraphics[width=\textwidth,height=0.82\textheight,keepaspectratio]{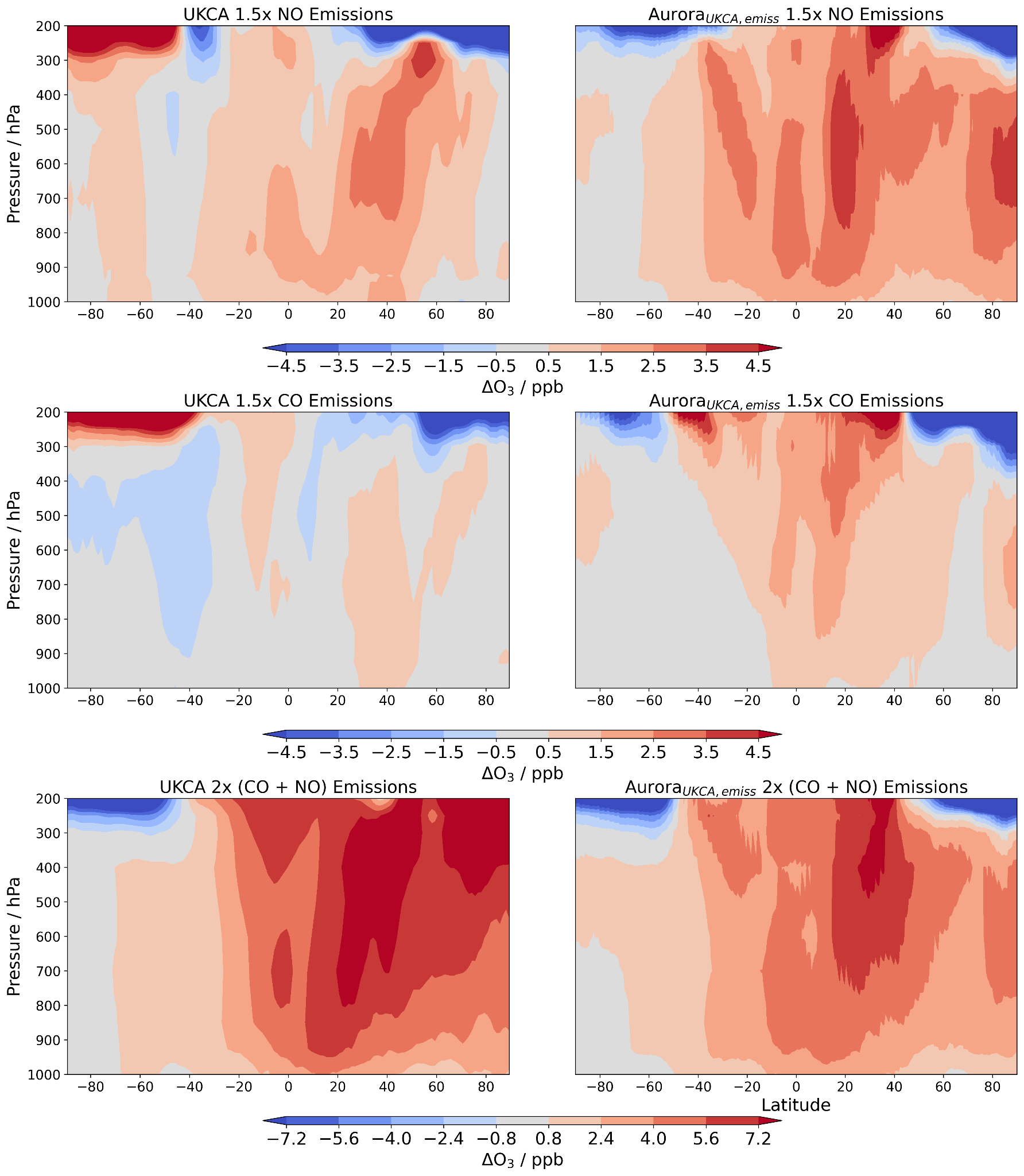}
\caption{\textbf{Zonal mean ozone responses to emission scenarios withheld from training.} Change in annual zonal mean ozone mole fraction relative to the control following scalings of 1.5 to NO emissions (top row), 1.5 to CO emissions (middle row), and 2 to both NO and CO emissions (bottom row).}
\label{fig:S15}
\end{figure}

\begin{figure}[htbp]
\centering
\includegraphics[width=\textwidth,height=0.82\textheight,keepaspectratio]{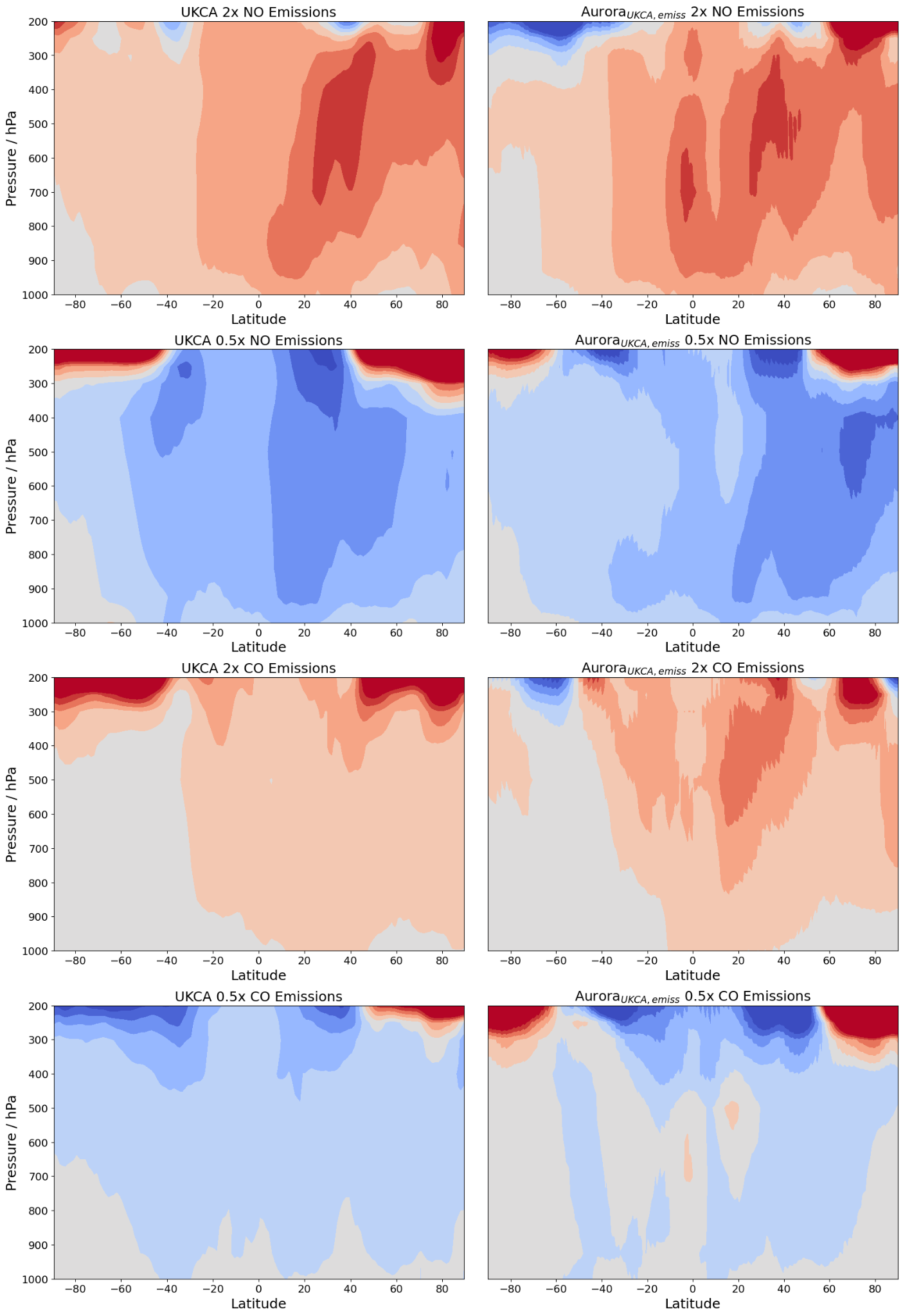}
\caption{\textbf{Zonal mean ozone responses to emission scalings used in training.} Change in annual zonal mean ozone mole fraction relative to the control following scalings of 2 to NO emissions (first row), 0.5 to NO emissions (second row), 2 to CO emissions (third row) and 0.5 to CO emissions (fourth row).}
\label{fig:S16}
\end{figure}

\begin{figure}[htbp]
\centering
\includegraphics[width=\textwidth,height=0.82\textheight,keepaspectratio]{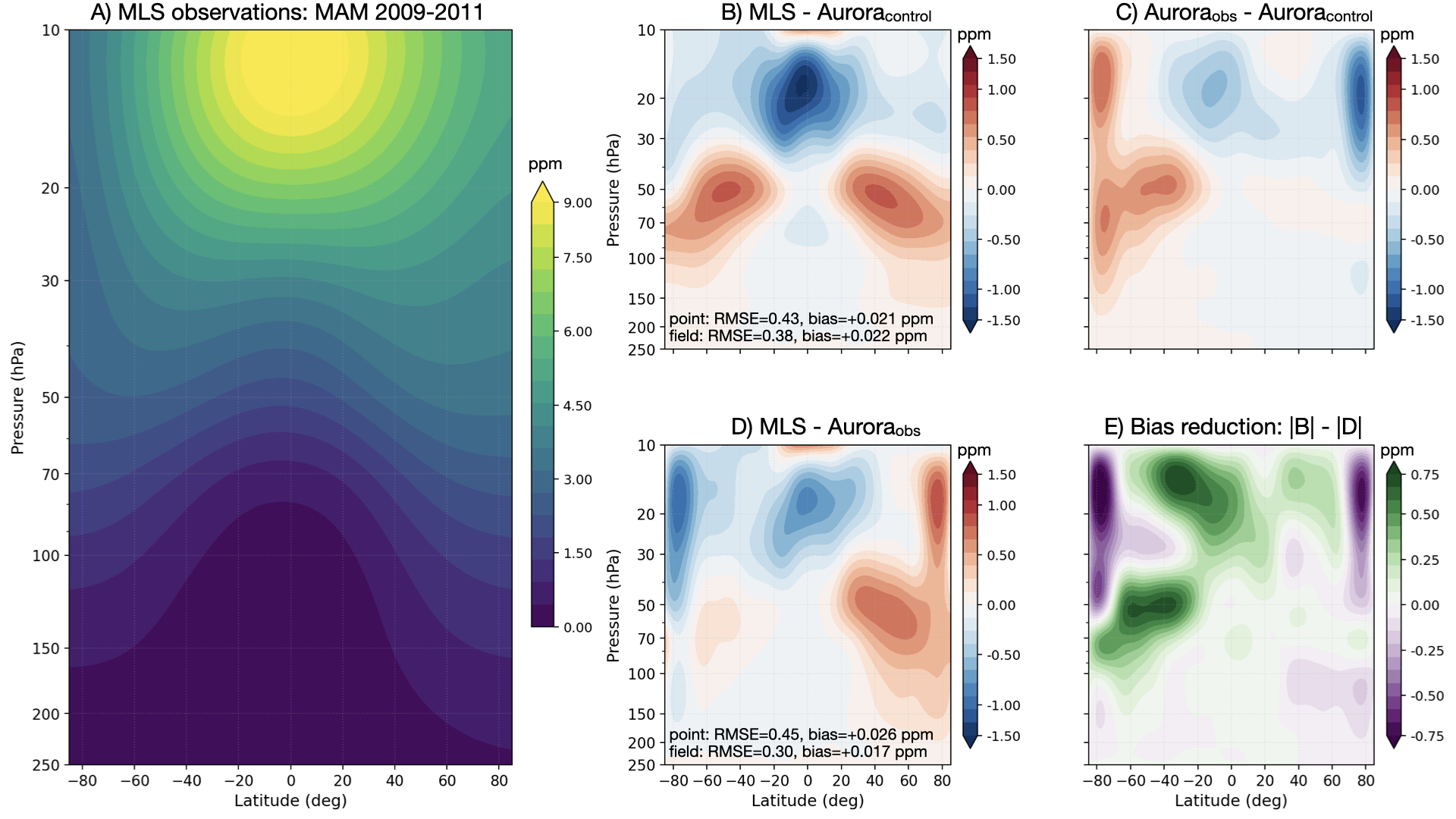}
\caption{\textbf{MLS-induced corrections as in Figure 6, but for MAM.}}
\label{fig:S17}
\end{figure}

\begin{figure}[htbp]
\centering
\includegraphics[width=\textwidth,height=0.82\textheight,keepaspectratio]{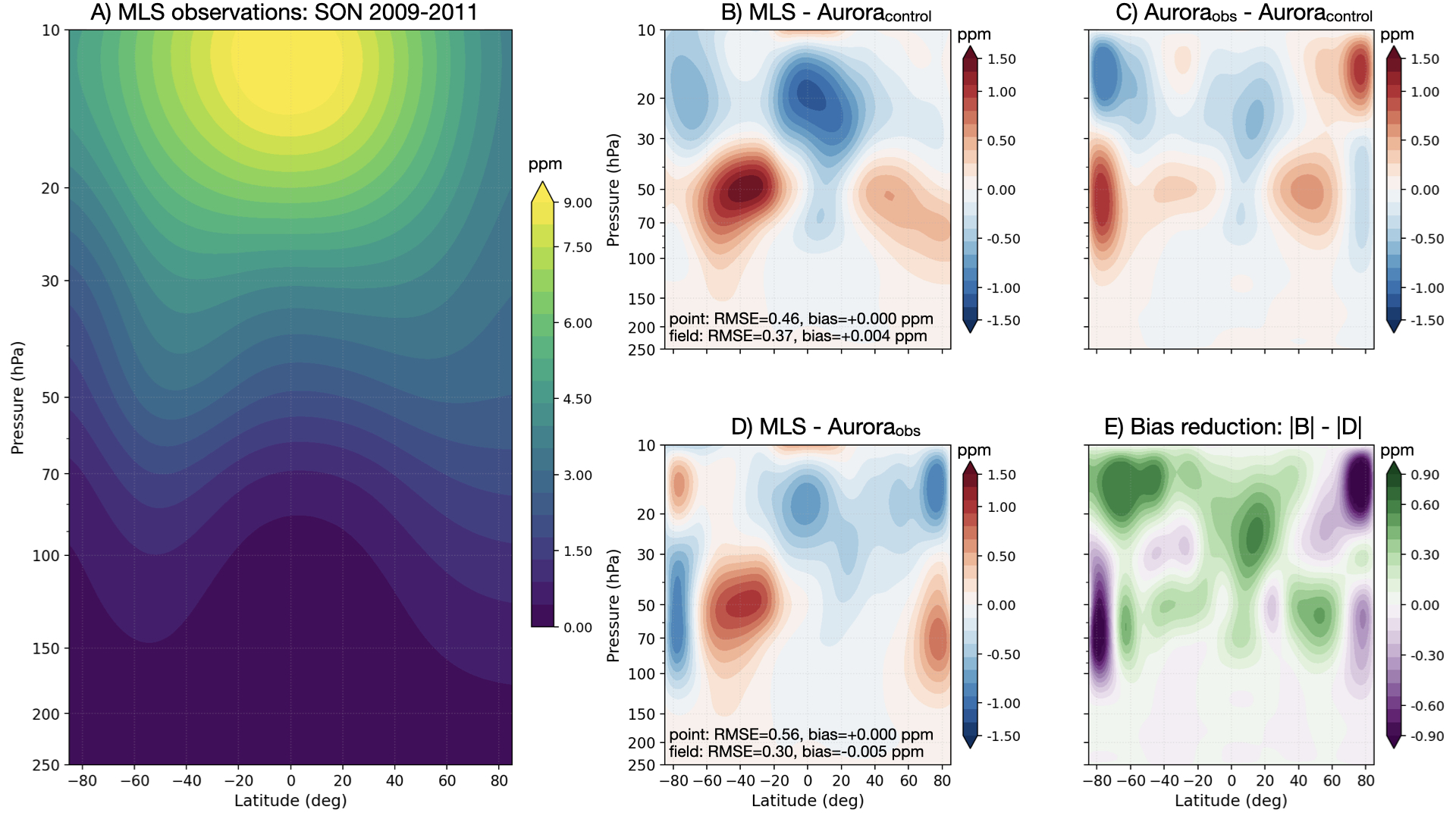}
\caption{\textbf{MLS-induced corrections as in Figure 6, but for SON.}}
\label{fig:S18}
\end{figure}

\begin{figure}[htbp]
\centering
\includegraphics[width=\textwidth,height=0.82\textheight,keepaspectratio]{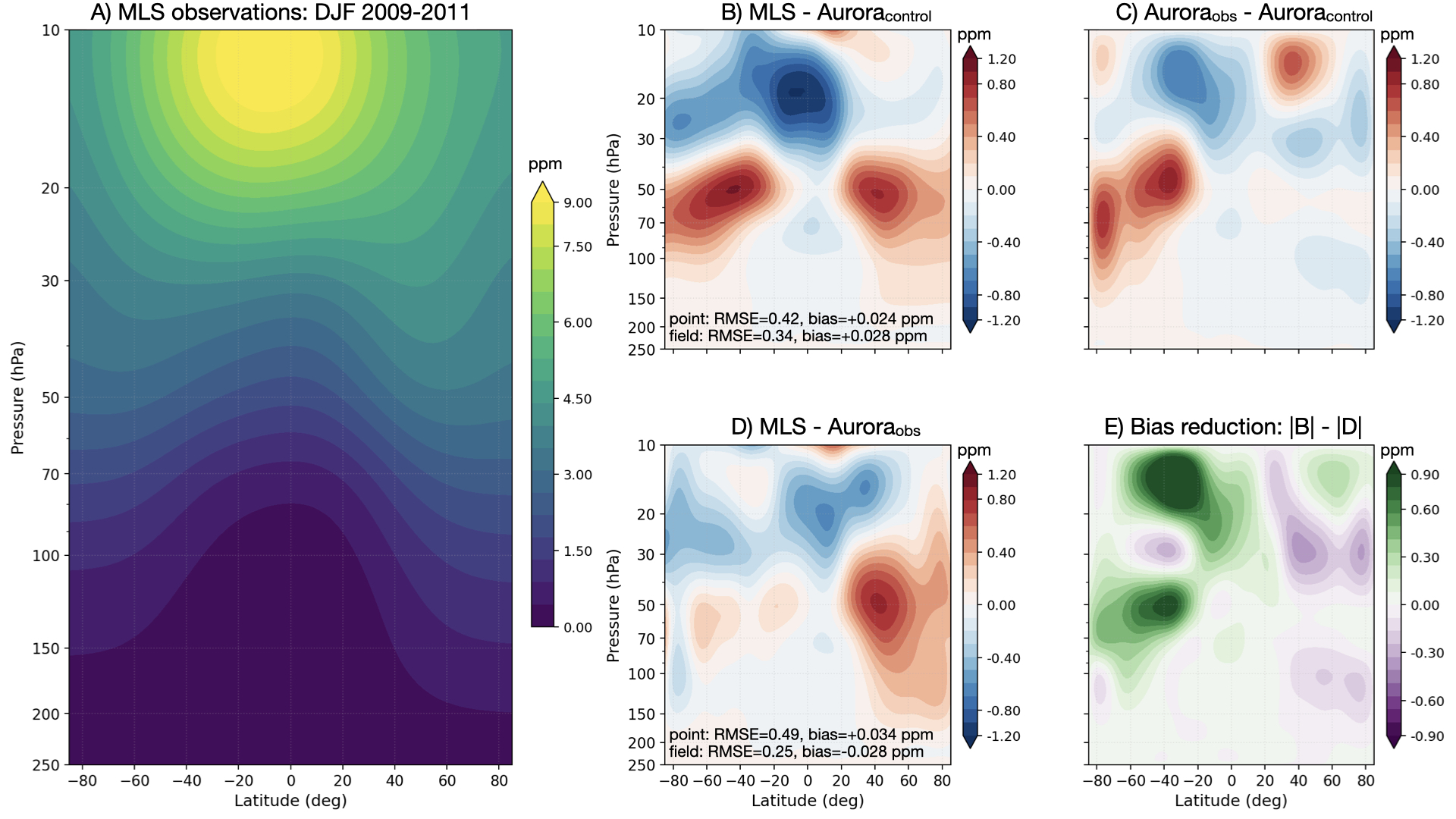}
\caption{\textbf{MLS-induced corrections as in Figure 6, but for DJF.}}
\label{fig:S19}
\end{figure}

\end{document}